\documentclass[aps,prb,preprint,showpacs,floatfix]{revtex4-1}

\usepackage{subcaption}
\usepackage{graphicx}
\usepackage{amssymb,amsfonts,amsmath}
\usepackage{dcolumn}

\def\bb{{\bf b}}
\def\bk{{\bf k}}
\def\br{{\bf r}}
\def\bR{{\bf R}}
\def\xh{\hat{x}}
\def\yh{\hat{y}}
\def\zh{\hat{z}}

\usepackage{color}

\begin{document}

\title{Band Structure Theory of the BCC to HCP Burgers Distortion}

\author{Bojun Feng and Michael Widom}
\affiliation{Department of Physics, Carnegie Mellon University, Pittsburgh, PA  15213}

\begin{abstract}
The Burgers distortion is a two-stage transition between body centered cubic (BCC) and hexagonal close-packed (HCP) structures. Refractory metal elements from the Sc and Ti columns of the periodic table (BCC/HCP elements) form BCC structures at high temperatures but transition to HCP at low temperatures via the Burgers distortion. Elements of the V and Cr columns, in contrast, remain BCC at all temperatures. The energy landscape of BCC/HCP elements exhibits an alternating slide instability, while the normal BCC elements remain stable as BCC structures. This instability is verified by the presence of unstable elastic constants and vibrational modes for BCC/HCP elements, while those elastic constants and modes are stable in BCC elements. We show that a pseudogap opening in the density of states at the Fermi level drives the Burgers distortion in BCC/HCP elements, suggesting the transition is of the Jahn-Teller-Peierls type. The pseudogap lies below the Fermi level for regular BCC elements in the V and Cr columns of the periodic table. The wave vector $\bk_S$ when the gap opens relates to the reciprocal lattice vector \textbf{G}=(1 $\frac{1}{2}$ $\frac{1}{2}$) of the distorted BCC structure as $\bk_S$=$\frac{1}{2}$\textbf{G}. BCC binary alloys containing both BCC/HCP and BCC elements exhibit a similar instability but stabilize part way through the BCC to HCP transition.
\end{abstract}

\maketitle

\section{Introduction}

Elements from the Sc and Ti columns of periodic table are body centered cubic (BCC) at high temperatures and transform to hexagonal close-packed (HCP) at low temperatures. We refer to these as BCC/HCP elements. Their transition is known as the Burgers distortion~\cite{burgers1934process}. On the other hand, elements from the V and Cr columns are BCC at all temperature below their melting temperatures. Although prior works discuss the BCC to HCP transition, (e.g. angular distortive matrices~\cite{cayron2016angular}, space group representation~\cite{srinivasan2002mechanism}, pressure-induced transitions~\cite{johnson2008nonadiabaticity, mankovsky2013pressure,chen1988first,wentzcovitch1988theoretical,craievich1997structural} and MD simulation~\cite{zhao2000thermal}), a complete understanding including the electronic structure driving the transition is missing. We seek the underlying cause of the instability in order to understand the mechanical properties of refractory metals and their alloys, especially high entropy alloys (HEAs) containing both BCC/HCP and normal BCC elements~\cite{feng2017elastic}. Our study is also complementary to the previous work of Lee and Hoffmann~\cite{lee2002bcc} who discussed a Jahn-Teller type transition of transition metals and alloys from BCC to FCC structures. While they focus on transition metals starting from V column going to the right of the periodic table, corresponding to the BCC to FCC transition, our study goes to the left of the periodic table. However, both transitions share similar pseudo-gap opening and bonding/antibonding orbital stabilization.

\begin{figure}
\begin{subfigure}{0.5\textwidth}
\includegraphics[trim = 8mm 25mm 20mm 30mm, clip, width=2.3in]{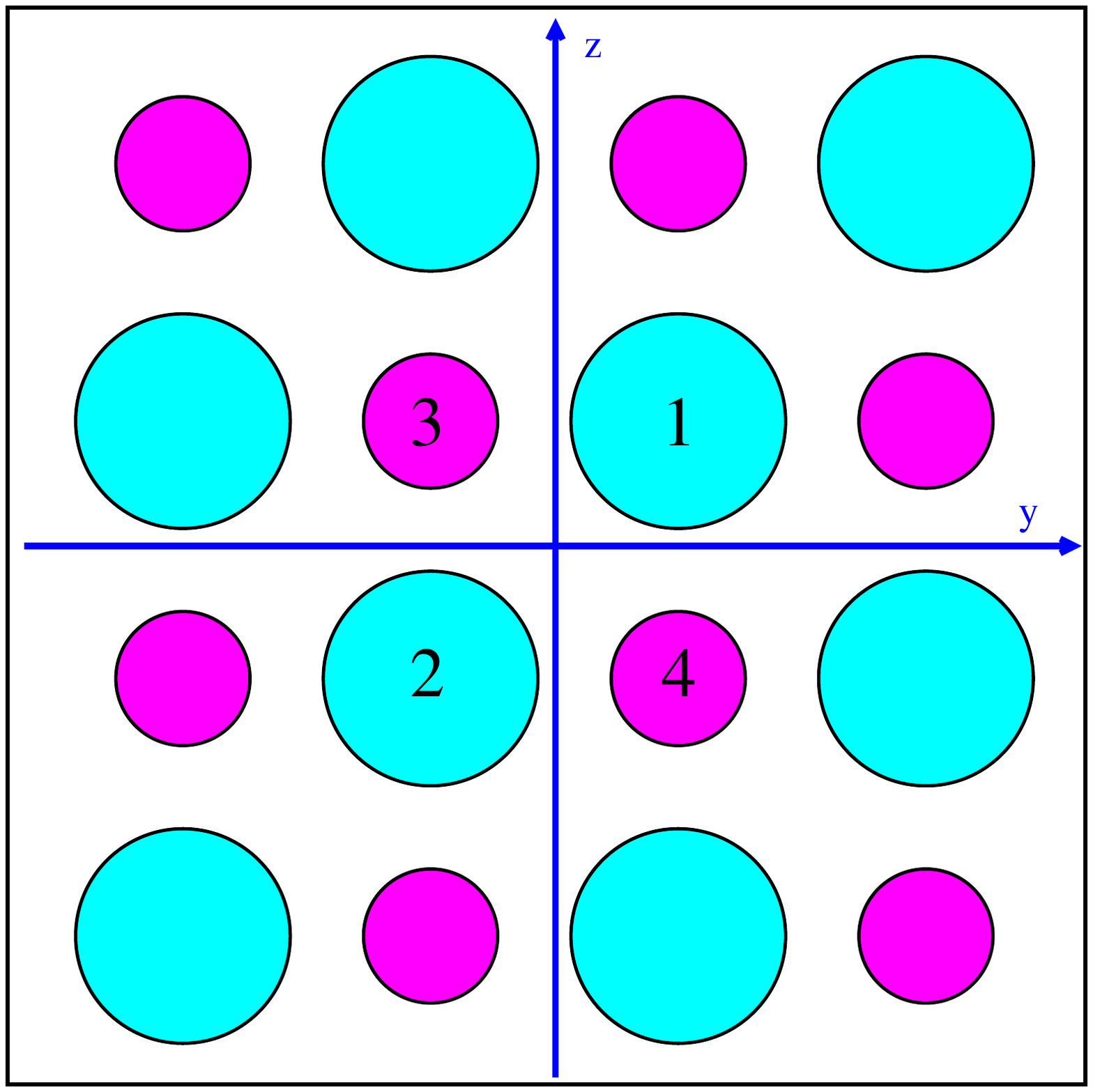}
\caption{$\lambda_1$=0, $\lambda_2$=0 (BCC)}
\label{fig:bsub00}
\end{subfigure}%
\begin{subfigure}{0.5\textwidth}
\includegraphics[trim = 8mm 25mm 20mm 30mm, clip, width=2.3in]{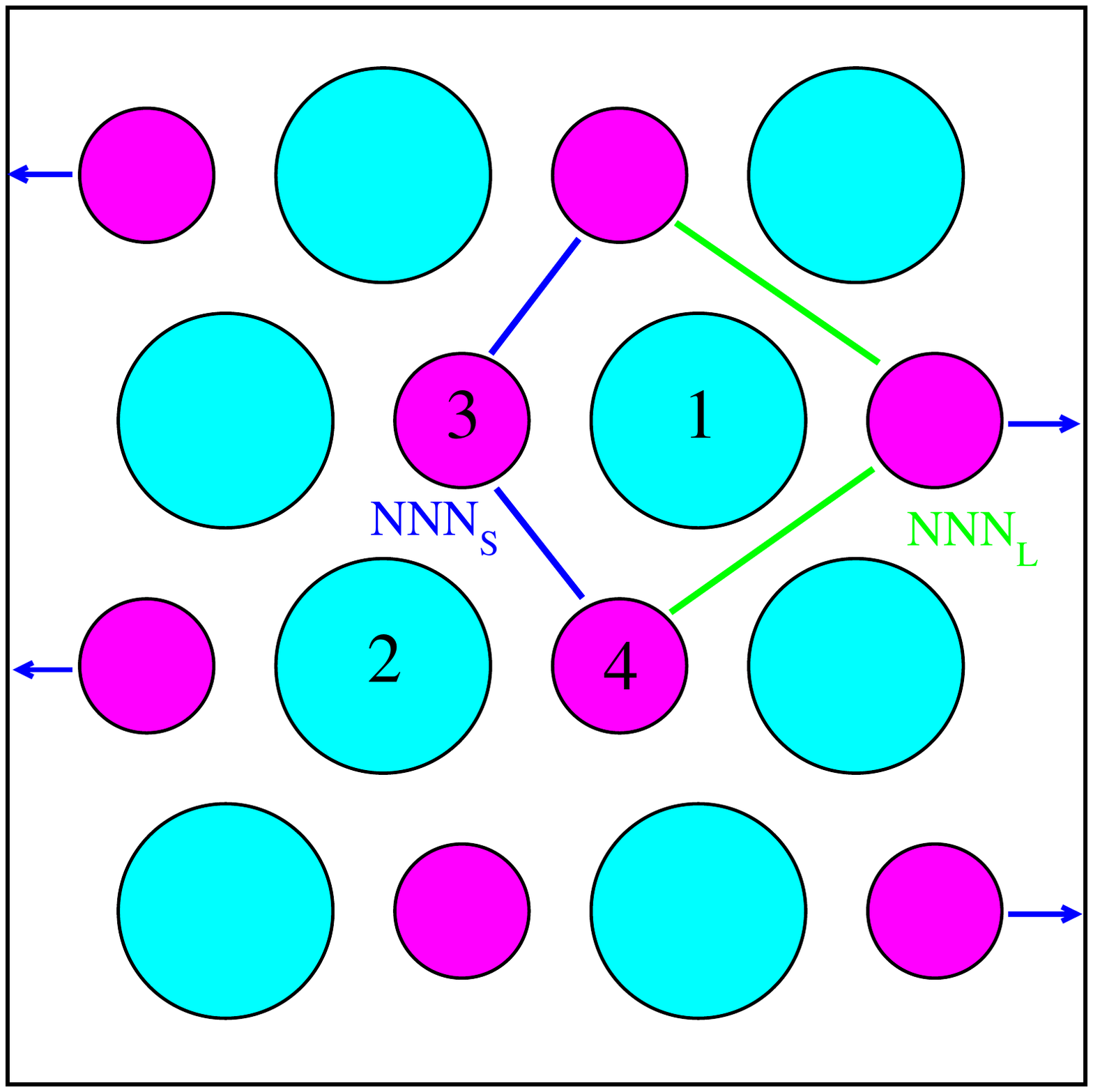}
\caption{$\lambda_1$=0, $\lambda_2$=1 (ORTHO)}
\label{fig:bsub10}
\end{subfigure}
\begin{subfigure}{0.5\textwidth}
\includegraphics[trim = 8mm 25mm 20mm 30mm, clip, width=2.3in]{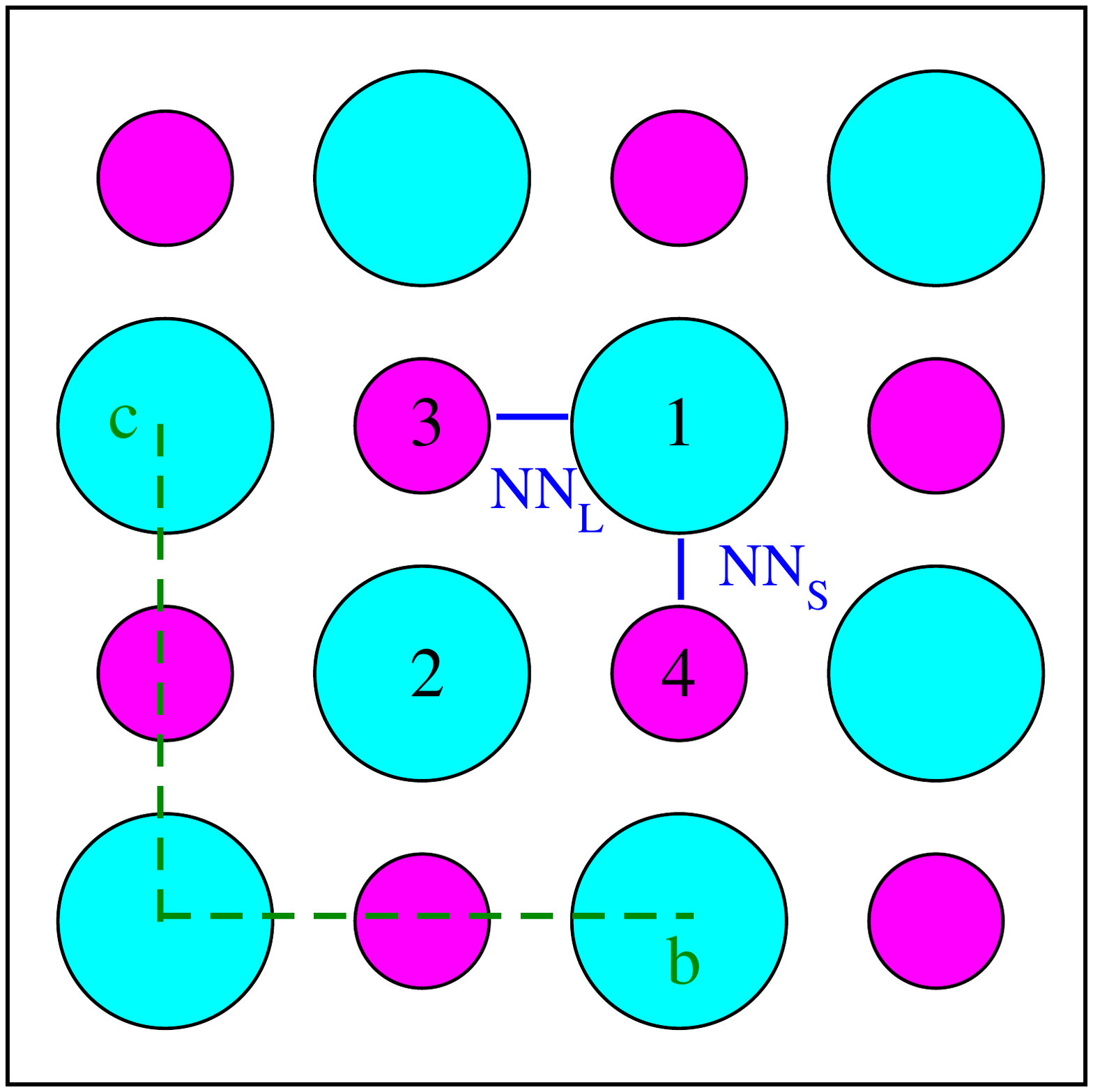}
\caption{$\lambda_1$=1, $\lambda_2$=0 (ORTHO)}
\label{fig:bsub01}
\end{subfigure}%
\begin{subfigure}{0.5\textwidth}
\includegraphics[trim = 8mm 25mm 20mm 30mm, clip, width=2.3in]{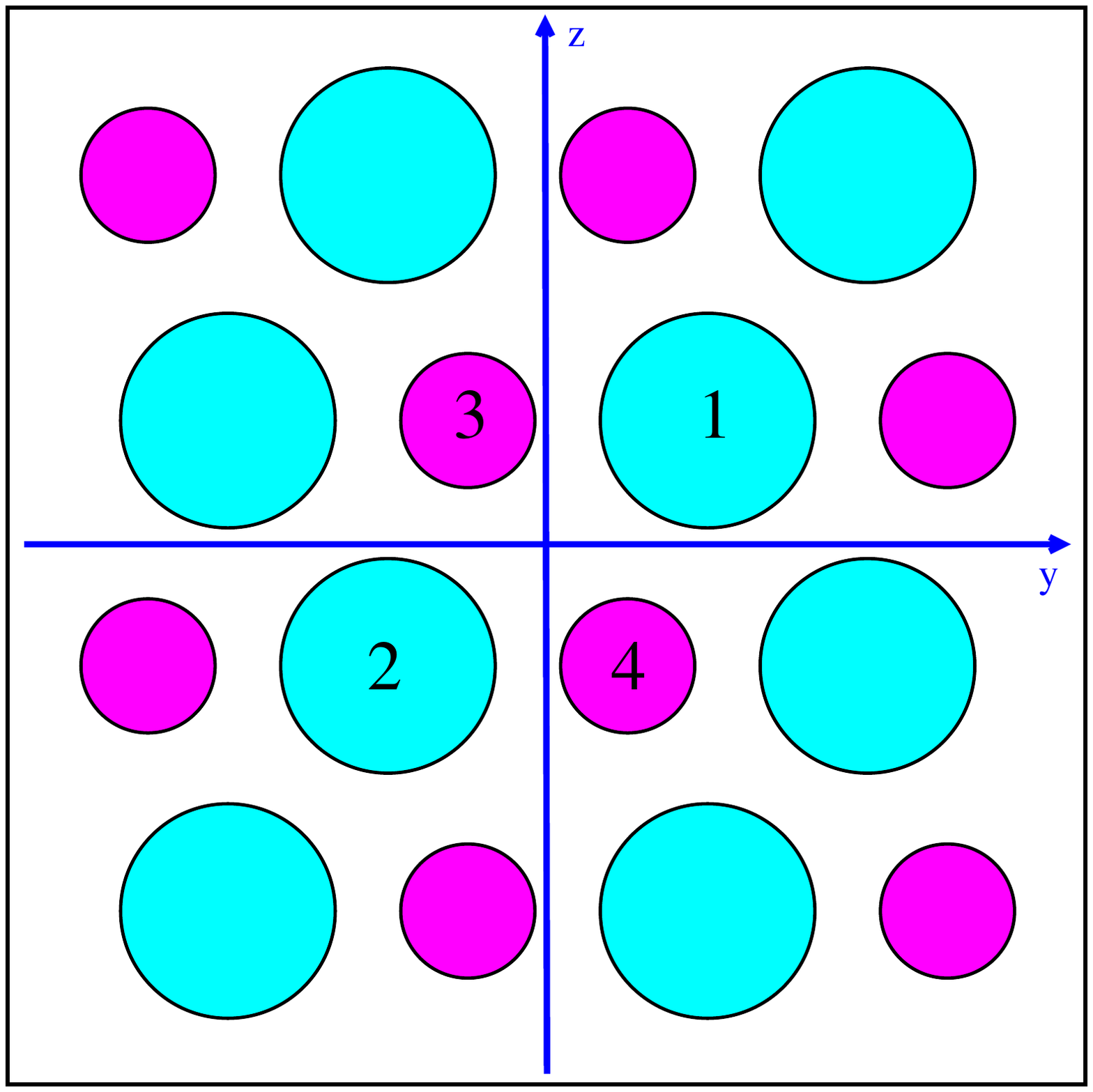}
\caption{$\lambda_1$=1, $\lambda_2$=1 (HCP)}
\label{fig:bsub111}
\end{subfigure}
\caption{Illustration of the Burgers distortion. (a) $\sqrt{2}\times\sqrt{2}\times1$ supercell of BCC viewed along the [1,0,0] direction; (b) Alternating slide displacement; (c) Orthorhombic shear b$>$c; (d) unit cell of HCP structure viewed along the [1,0,0] direction. Large cyan atoms are in lower layer; small magenta atoms are in upper layer. For pure element all atoms are same species; for binary alloys, magenta and cyan are BCC/HCP and normal BCC, respectively. Atoms are labeled according to Eq.~\ref{eq:R1234}.}
\label{fig:burgers}
\end{figure}

The Burgers distortion is a two-stage transition, consisting of an orthorhombic shear deformation and an alternating slide displacement between atomic layers of the BCC structure~\cite{masuda2004hcp}. We sketch the mechanism in Fig.~\ref{fig:burgers}. In the notation of Ref.~\cite{masuda2004hcp}, a Pearson type oS4 cell characterized by two variables, $\lambda_1$ and $\lambda_2$, interpolates between the BCC and HCP structures. The lattice constants of this oS4 cell are
\begin{equation}
\label{eq:abc}
a (\lambda_1) = a_0/\alpha(\lambda_1),~~
b (\lambda_1) = \alpha(\lambda_1)\sqrt{2}a_0,~~
c = \sqrt{2}a_0
\end{equation}
where $a$, $b$ and $c$ are the three lattice constants of the oS4 cell, $\alpha(\lambda_1)=1+(\sqrt[4]{3/2}-1)\lambda_1$, and $a_0$ is the lattice constant of the corresponding BCC structure. Notice that the lattice constants of the oS4 cell only depend on the value of $\lambda_1$ and their variation generates orthorhombic shear. The positions of the four atoms in the oS4 cell are
\begin{align}
\label{eq:R1234}
\bR_1 &= (0, \frac{3+\lambda_2}{12}b, \frac{1}{4}c ) &
\bR_2 &= (0, -\frac{3+\lambda_2}{12}b, -\frac{1}{4}c ) \nonumber \\
\bR_3 &= (\frac{1}{2}a, \frac{\lambda_2 - 3}{12}b, \frac{1}{4}c ) &
\bR_4 &= (\frac{1}{2}a, \frac{3-\lambda_2}{12}b, -\frac{1}{4}c ).
\end{align}
$\lambda_2$ generates the alternating slide between atomic layers in (1,1,0) planes of BCC. When $\lambda_1=\lambda_2$=0, we have a BCC structure (Pearson type cI2), and the oS4 cell is a 1$\times\sqrt{2}\times\sqrt{2}$ supercell of BCC. When $\lambda_1=\lambda_2$=1, the structure is HCP (Pearson type hP2).

Both $\lambda_1$ and $\lambda_2$ alter bond bond lengths. Nearest neighbor (NN) and next nearest neighbor (NNN) bonds are of particular importance. To understand the instability of the BCC structure, we consider their variation for small distortions. Expanding to first order in $\lambda_1$ and $\lambda_2$ we find
\begin{eqnarray}
\label{eq:expand}
NN_L &\equiv |\bR_1-\bR_3| &\approx
\frac{\sqrt{3}}{2}a_0+\frac{1}{12}(6^{3/4}-2\sqrt{3})a_0\lambda_1 \\ \nonumber
NN_S &\equiv |\bR_1-\bR_4| &\approx
\frac{\sqrt{3}}{2}a_0-\frac{1}{12}(6^{3/4}-2\sqrt{3})a_0\lambda_1 \\ \nonumber
NNN_L &\equiv |\bR_1-\bR_2| &\approx a_0+\frac{1}{6}a_0\lambda_2 \\ \nonumber
NNN_S &\equiv |\bR_3-\bR_4| &\approx a_0-\frac{1}{6}a_0\lambda_2 \\ \nonumber
\end{eqnarray}
Notice that both the NN and NNN bonds split into a long and a short version, with the NN bonds varying to first order only in $\lambda_1$, and the NNN bonds varying to first order only in $\lambda_2$.

In the following sections, we examine the impact of electronic structure on the total energy as $\lambda_1$ and $\lambda_2$ vary during the Burgers distortion. We show that $\lambda_2$ distortion drives the initial instability of the BCC structure, by opening a gap in the electronic band structure, creating a pseudo-gap in the electronic density of state and a charge density wave, with subsequent relaxation in the $\lambda_1$ variable that eventually stabilizes an HCP structure. We recognize the initial instability as a type of Jahn-Teller-Peierls distortion~\cite{Jahn1937,Peierls1955,englman1972jahn, burdett1983peierls}. We then turn to alloys and show how the BCC structure of binary alloys containing both BCC/HCP and normal BCC elements are stabilized part way through Burgers the distortion.

\section{Pure Elements}
\subsection{Elasticity and Phonons}
\subsubsection{Elasticity}
We begin our analysis with calculation of BCC/HCP and normal BCC refractory element T=0K elastic constants. These are obtained within density functional theory from stress-strain relationships using two-point central differences as implemented in VASP~\cite{kresse1999ultrasoft}. We employ the generalized gradient approximation~\cite{perdew1996generalized} without spin polarization. The energy cutoffs of the plane wave basis sets are set to 400eV and k-point meshes are set to 14$\times$14$\times$14 in 16-atom 2$\times$2$\times$2 supercells of the conventional 2-atom unit cell. We use ``Accurate'' precision to avoid wraparound errors. 

As shown in Table ~\ref{tab:elasticp}, BCC/HCP elements from the Sc and Ti columns have $C_{11}\leq C_{12}$. This violates a Born stability condition and predicts instability to a tetragonal or orthorhombic distortion. These elements are stabilized in the BCC state at high temperature by their vibrational entropy~\cite{souvatzis2008entropy}. The BCC structures become mechanically unstable at low temperatures, causing the transformation to HCP. Elements from the V and Cr columns all have $C_{11}>C_{12}$, so that the BCC structures are maintained at low temperatures.

\begin{table}
\caption{\label{tab:elasticp} Calculated T=0K elastic constants of elements from Sc-Cr columns of the periodic table (units of Gpa).}
\begin{tabular}{c|ccc|ccc|ccc|ccc}
\toprule
moduli   & C$_{11}$ & C$_{12}$ & C$_{44}$ & C$_{11}$  & C$_{12}$ & C$_{44}$  & C$_{11}$ & C$_{12}$ & C$_{44}$ & C$_{11}$  & C$_{12}$ & C$_{44}$ \\
\hline
\hline
element &         &    Sc   &         &         & Ti      &         &          &    V     &          &          &  Cr      &         \\
\hline
moduli   &   59    &   59     &   27     &    99   &  119    &   41    &   317    &   163    &  28   &   580    &    175   &     119   \\
\hline
\hline
element &         &    Y     &         &         & Zr      &         &          &    Nb     &        &          &    Mo    &           \\
\hline
moduli   &   25    &    46    &    22    &   92    &   96    &   34    &    250   &   139    &  17    &     517  &    181   &   117     \\
\hline
\hline
element &         &     La   &          &         & Hf      &         &          &    Ta     &       &          &     W    &           \\
\hline
moduli   &    -14  &     47   &     7    &   77    &  118    &    54   &     270  &    163   &  77    &     525  &    205   &   147    \\
\end{tabular}
\end{table}

\subsubsection{Phonon Instability}
\begin{figure}
\includegraphics[width=3in]{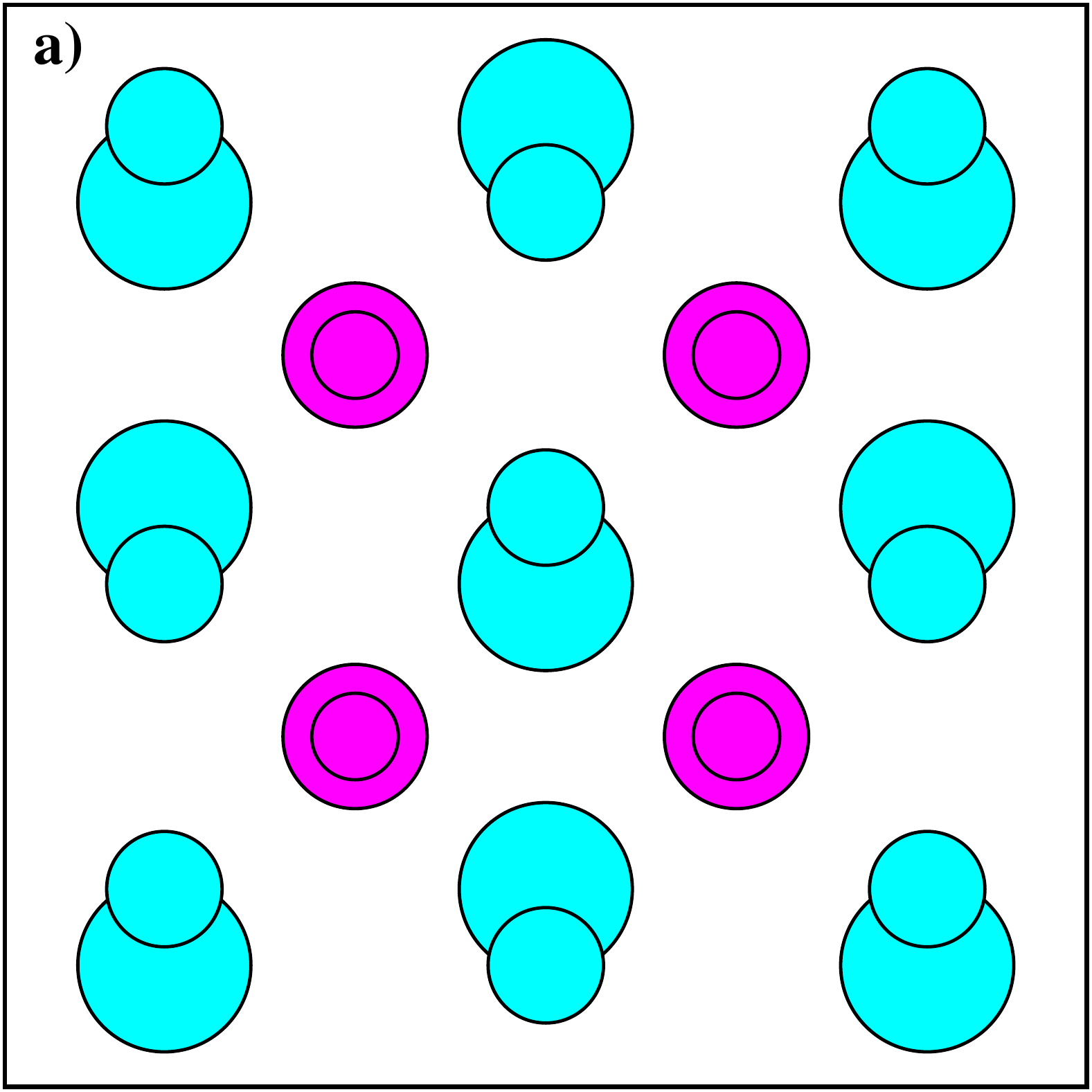}
\includegraphics[width=3in]{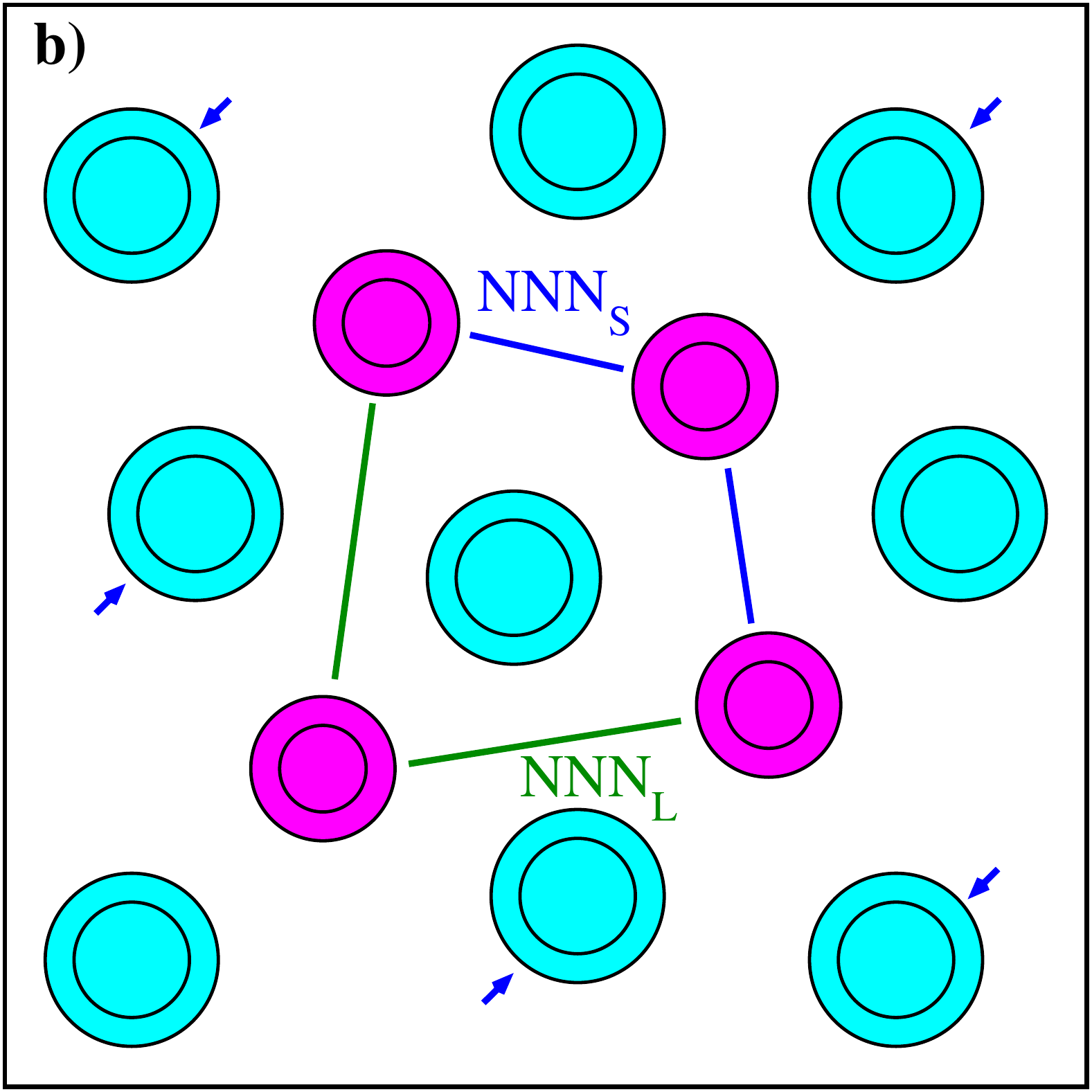}
\caption{\label{fig:modes} Unstable phonon modes of Hf in $2\times 2\times 2$ supercell of conventional BCC unit cell viewed along the cubic [100] axis. Note the $y$ and $z$ axes are rotated by 45$^{\circ}$ compared with Fig.~\ref{fig:burgers}. Atom sizes indicate vertical height. Colors distinguish cube vertex (magenta) from cube center (cyan). (a) Lower imaginary frequency. (b) Upper imaginary frequency.}
\end{figure}

We calculate the $\Gamma$-point phonon modes of the 2$\times$2$\times$2 supercell structures using density functional perturbation theory (DFPT)~\cite{baroni2001phonons}. Fig.~\ref{fig:modes} illustrates the unstable modes of Hf and Table~\ref{tab:phononp} lists the unstable phonon mode frequencies for all HCP/BCC elements. Each of the BCC/HCP elements has a 6-fold degenerate imaginary frequency mode, and three of them also have a second 6-fold degenerate lower imaginary frequency mode. All BCC elements are stable as BCC structures with no imaginary frequency modes. In every case the (maximal) imaginary frequency mode corresponds to the $\lambda_2$ alternating slide deformation illustrated in Fig.~\ref{fig:burgers}b. We can understand the 6-fold degeneracy because we have three choices for the direction of alternation ({\em i.e.} $\xh, \yh$, or $\zh$ in Fig.~\ref{fig:burgers}) and for each direction of alternation we have two choices of perpendicular direction in which to displace. Equivalently, the cubic crystal system has 6 independent but symmetry equivalent \{111\} planes within which to slide. This mode reduces the symmetry from cubic to orthorhombic. If the initial cubic structure is displaced according to this mode, it follows the Burgers distortion pathway and relaxes to HCP. The lower imaginary frequency mode corresponds to a tetragonal symmetry breaking. If the initial cubic structure is displaced according to this mode it follows the Bain path to either a tI2 or an FCC structure.

\typeout{Before fig:modes}

\typeout{After fig:modes}

\begin{table}
\caption{\label{tab:phononp} The unstable (imaginary frequency) phonon modes of elements from Ti and V columns of periodic table (unit of frequency: THz, d$_{f_i}$: degeneracy of mode f$_i$).}
\begin{tabular}{ll|ll}
\toprule
element & frequency & element & frequency \\
\hline
\hline
Sc & 2.73$i$          & Ti & 3.21$i$, 4.93$i$ \\
Y  & 2.05$i$          & Zr & 2.50$i$ \\
La & 1.76$i$, 1.80$i$ & Hf & 1.80$i$, 2.76$i$
\end{tabular}
\end{table}

\subsubsection{Energy Landscape}
\begin{figure}
\begin{subfigure}{0.5\textwidth}
\includegraphics[trim = 0mm 60mm 0mm 60mm, clip, width=2.7in]{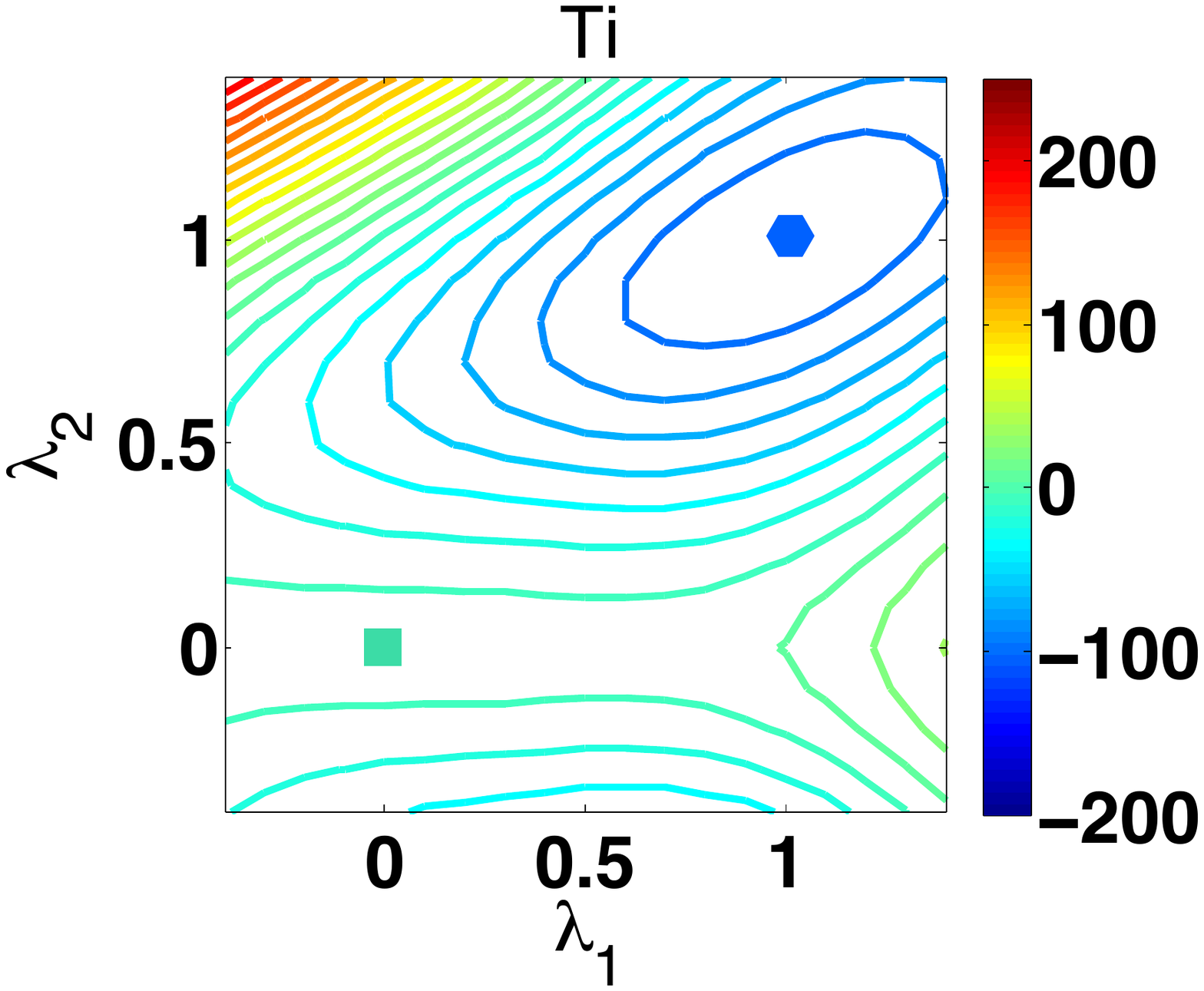}
\label{fig:tie}
\end{subfigure}%
\begin{subfigure}{0.5\textwidth}
\includegraphics[trim = 0mm 60mm 0mm 60mm, clip, width=2.7in]{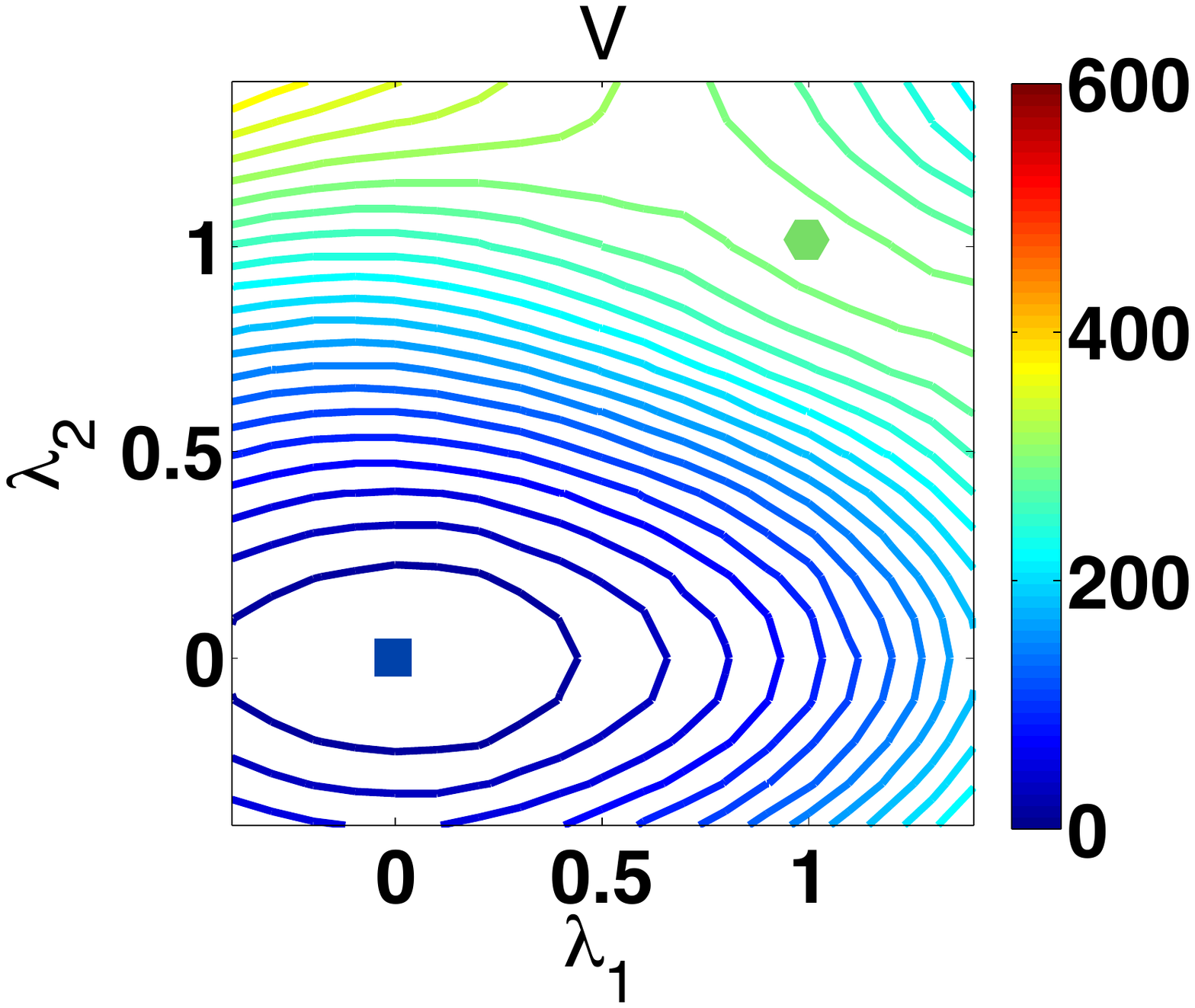}
\label{fig:ve}
\end{subfigure}
\begin{subfigure}{0.5\textwidth}
\includegraphics[trim = 0mm 60mm 0mm 60mm, clip, width=2.7in]{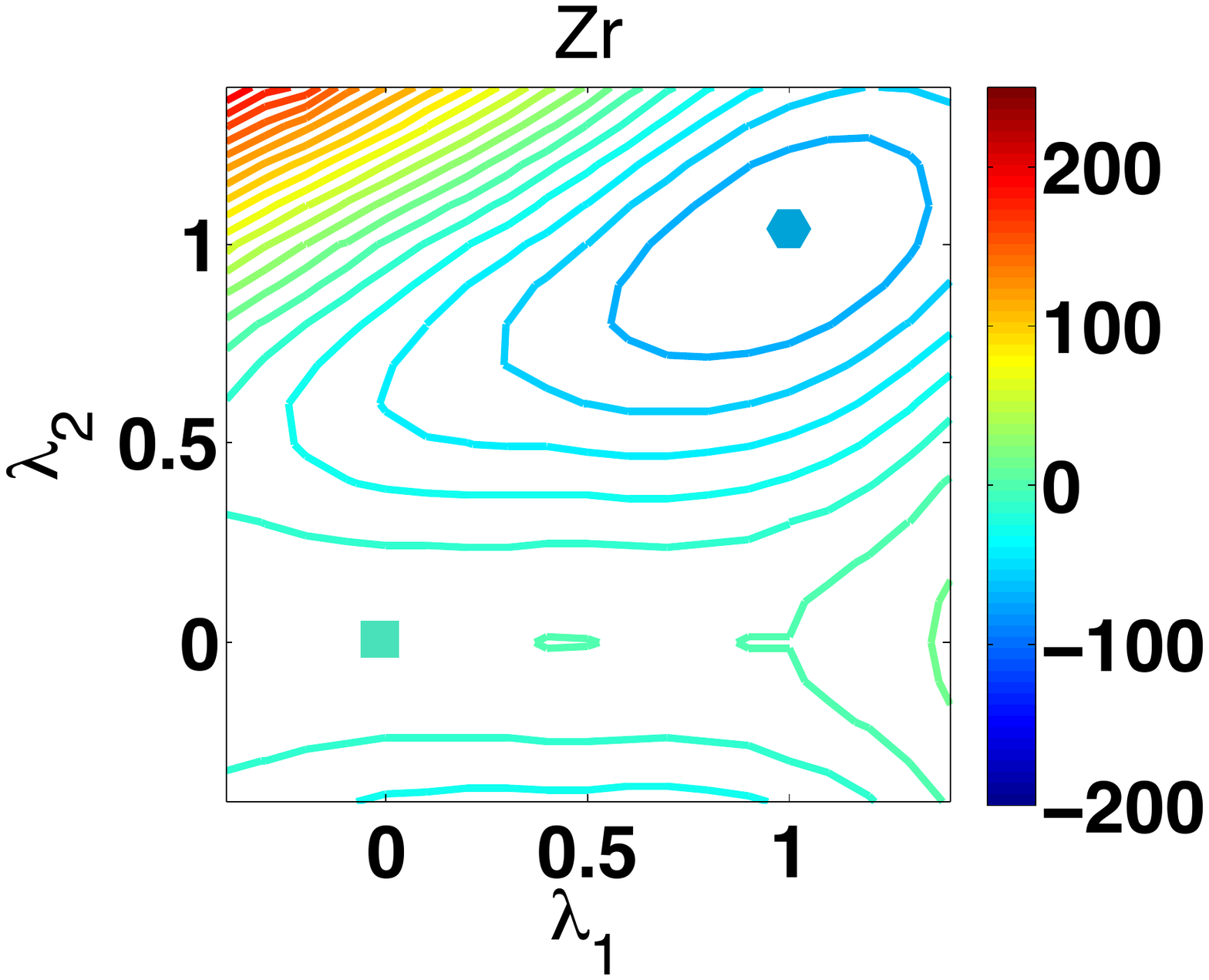}
\label{fig:zre}
\end{subfigure}%
\begin{subfigure}{0.5\textwidth}
\includegraphics[trim = 0mm 60mm 0mm 60mm, clip, width=2.6in]{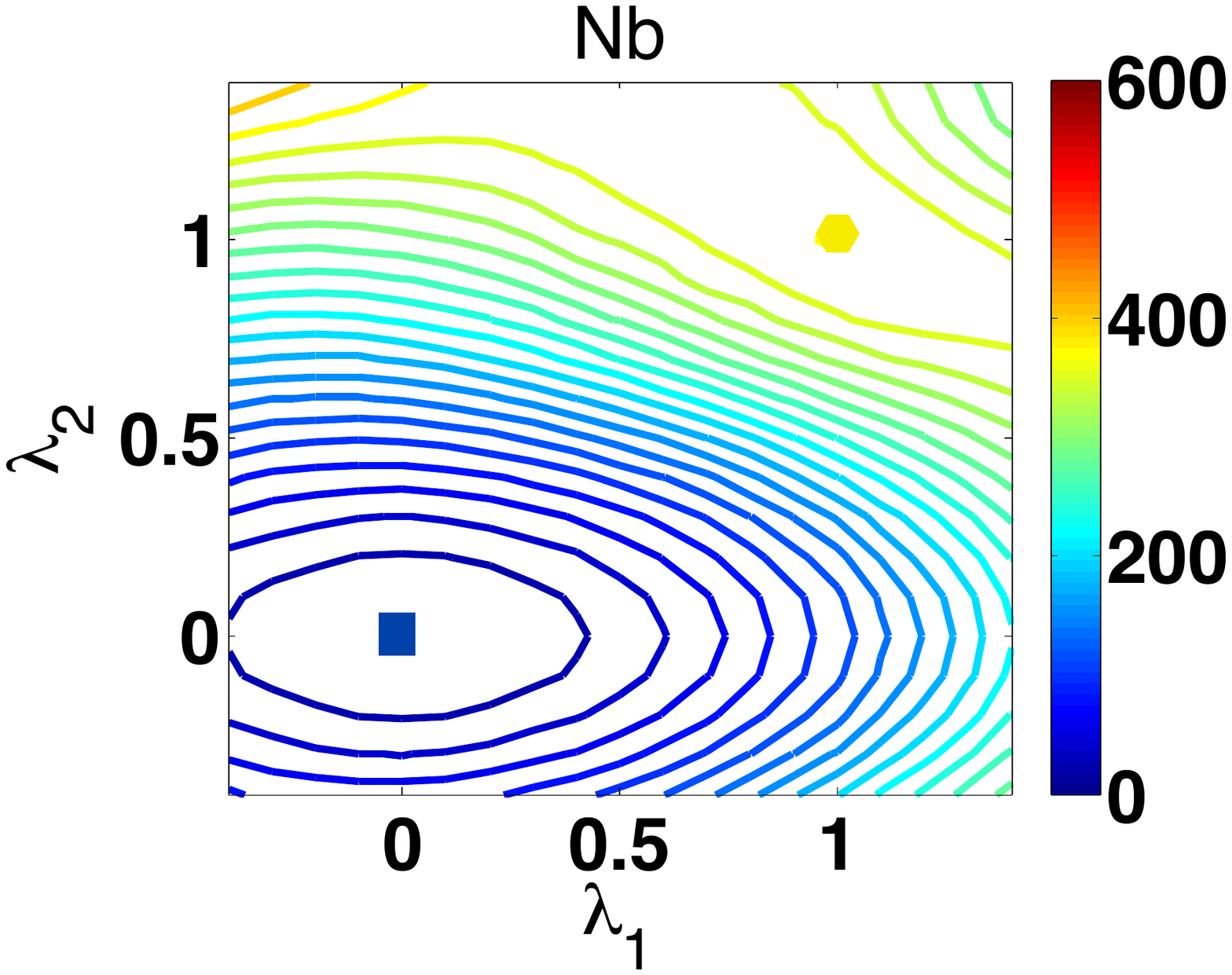}
\label{fig:nbe}
\end{subfigure}
\begin{subfigure}{0.5\textwidth}
\includegraphics[trim = 0mm 60mm 0mm 60mm, clip, width=2.7in]{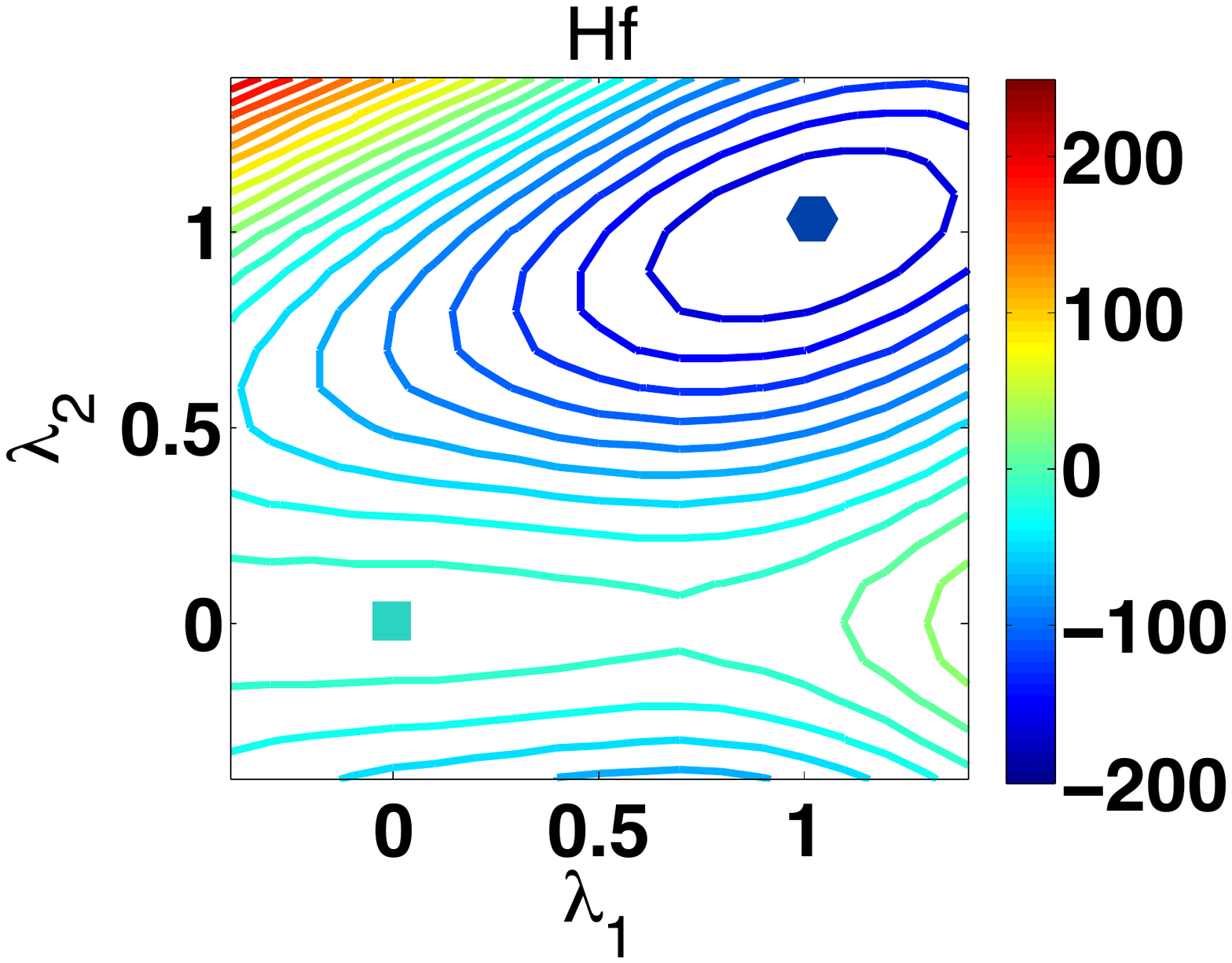}
\label{fig:hfe}
\end{subfigure}%
\begin{subfigure}{0.5\textwidth}
\includegraphics[trim = 0mm 60mm 0mm 60mm, clip, width=2.7in]{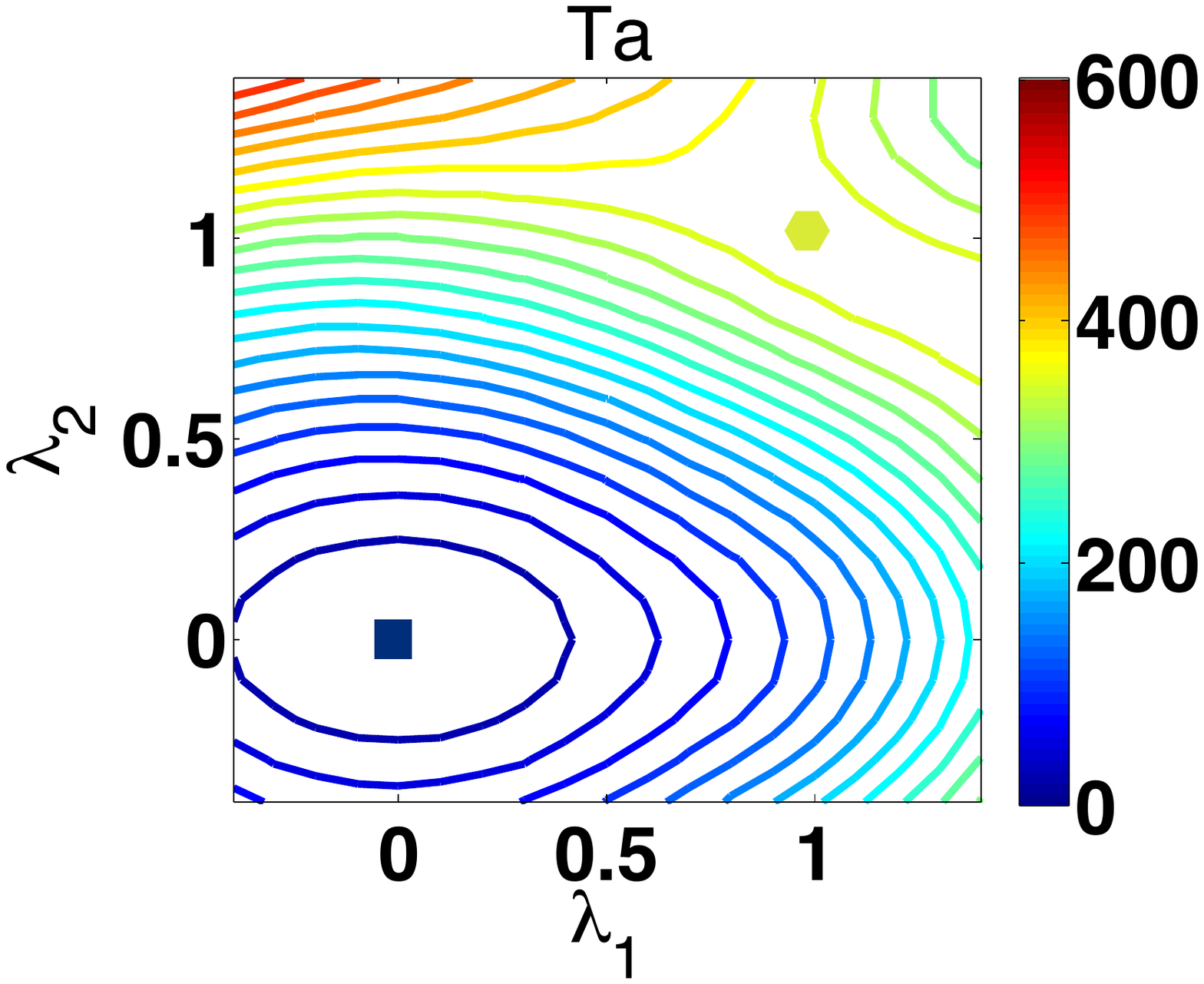}
\label{fig:tae}
\end{subfigure}
\caption{Energy landscapes of elements from the Ti (left) and V (right) columns of the periodic table. Elements from the Ti column are stable as HCP (hexagons), whereas elements from the V column are stable as BCC (squares). Color bars give energy contours in meV/atom relative to energies of the BCC structures.}
\label{fig:elp}
\end{figure}

The instabilities of the BCC/HCP elements can be seen from their energy landscapes as the $\lambda_1$ and $\lambda_2$ values are varied (see Fig.~\ref{fig:elp}). These are calculated within the conventional oS4 unit cell using 12$\times$8$\times$8 $k$-point meshes and otherwise normal VASP defaults. Specifically, only lattice volume is relaxed, but not cell shape or ion position, in order to maintain the $\lambda_1$ and $\lambda_2$ values. BCC/HCP elements are more stable as HCP structures, while BCC elements are more stable as BCC structures. Notice that the Burgers distortion is driven initially by the $\lambda_2$ distortion, since if we start from BCC ($\lambda_1=0$, $\lambda_2=0$), changing $\lambda_2$ reduces the energy much more quickly than changing $\lambda_1$ does. Thus the Burgers distortion should begin in the $\lambda_2$ direction (alternating slide), then later complete in the $\lambda_1$ direction (variation of lattice parameters).

Fig.~\ref{fig:elp} presents energy landscapes for the Ti and V columns of the periodic table. Similar behaviors are found in the Sc and Cr columns, although in the case of the Sc column the initial instabilities in $\lambda_1$ are somewhat stronger than those in the Ti column.

\subsection{Electronic Structure}
So far, our investigation of elasticity, phonon modes and energy landscapes has illustrated the instability of BCC/HCP elements without revealing the underlying mechanism. Here, we seek an explanation by examining the electronic structure. Our study focuses on the HCP/BCC element Hf and the normal BCC element Ta. Our findings for Ta apply equally to the entire V and Cr columns of the periodic table, while our findings for Hf apply to the entire Ti column, and with minor modification (discussed later), to the Sc column.

\subsubsection{Density of States}

\begin{figure}
\includegraphics[width=3in]{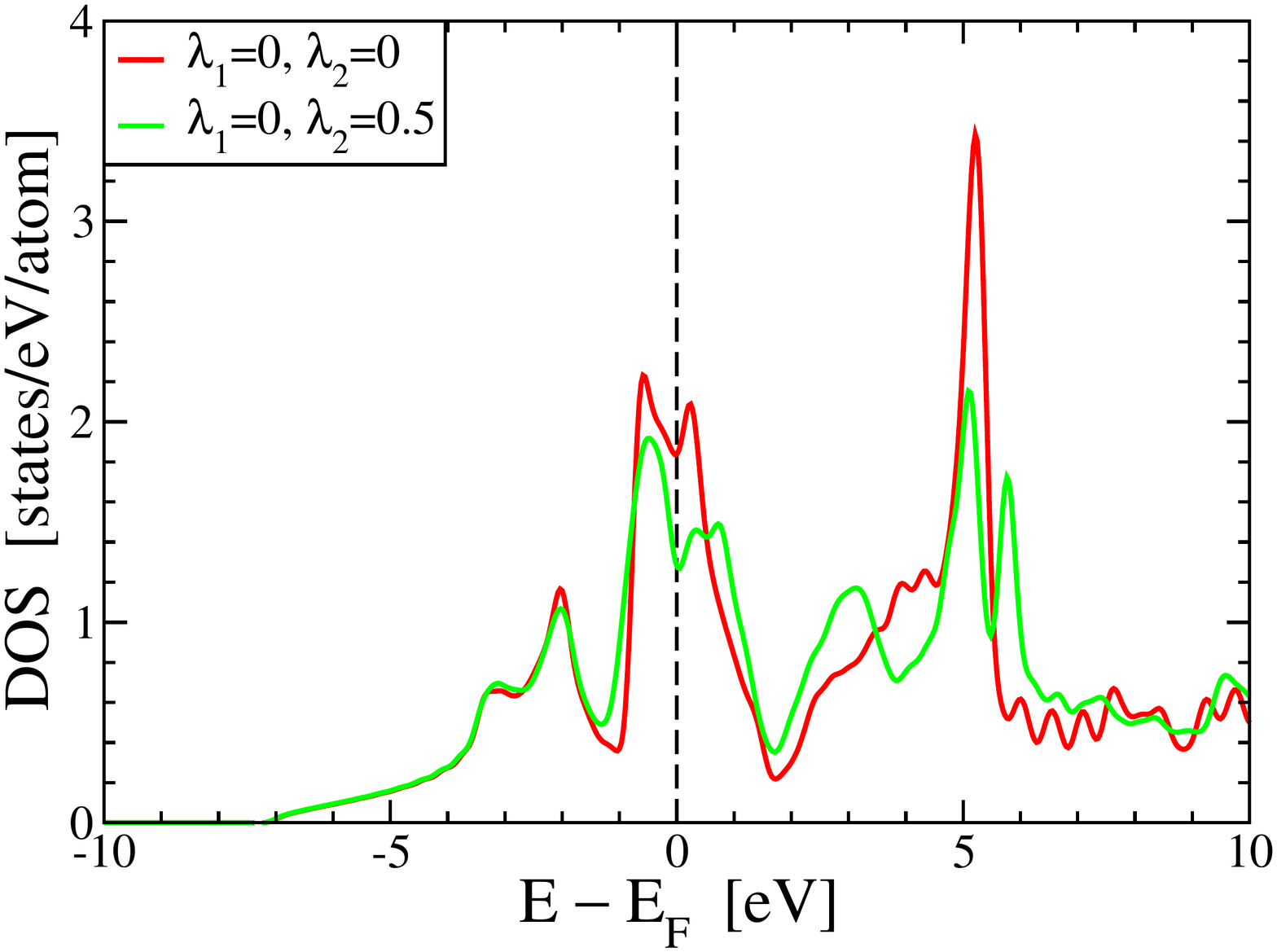}
\includegraphics[width=3in]{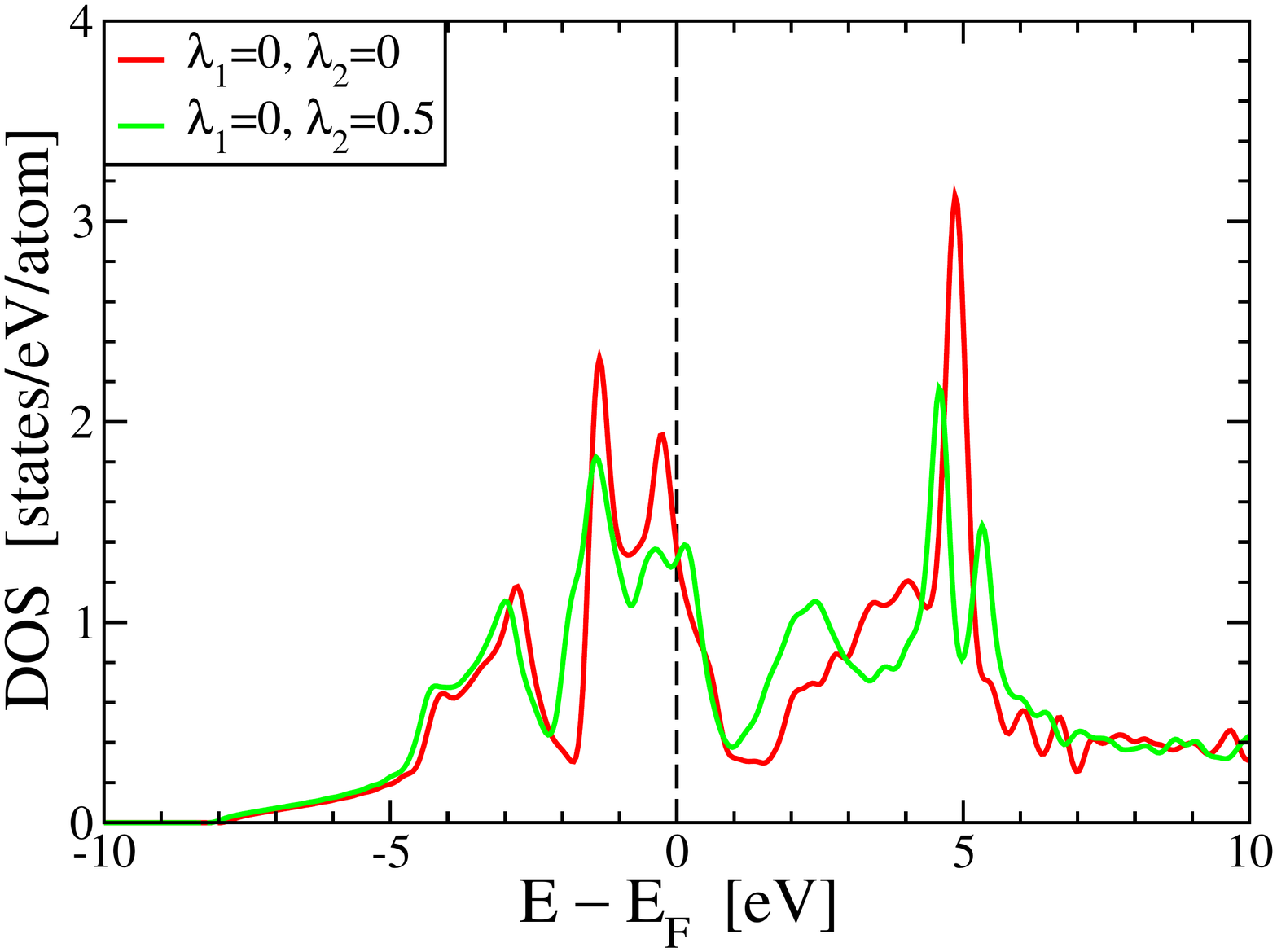}
\caption{DOS comparison of Hf (left) and Ta (right) before and after the application of $\lambda_2$ distortion.}
\label{fig:dosp}
\end{figure}

The electronic density of states (DOS) is qualitatively similar for Hf and Ta, although the Fermi energy, $E_F$, is higher for Ta owing to its extra valence electron. Hf has a weak pseudogap right at $E_F$, while this pseudogap lies below $E_F$ in the case of Ta. The pseudogap deepens upon application of the $\lambda_2$ distortion, as illustrated in Fig.~\ref{fig:dosp}. Thus, in the case of Hf, increasing $\lambda_2$ reduces the energy of occupied states below $E_F$ while raising the energy of empty states above $E_F$, and hence lowering the band energy~\cite{hume1926researches, jones1934theory, jones1962theory} relative to the initial BCC structure~\cite{Fujiwara1991,de2005electronic,Mizutani2010}. The band energy of Ta is less strongly affected, because the pseudogap opening occurs below $E_F$.

The impact of $\lambda_2$ on total energy $E_{Total}(\lambda_2)$ is quantified in Table~\ref{tab:epartp} for Hf and Ta. Here $E_0$ refers to values at $\lambda_2=0$. Owing to the symmetry between $\pm \lambda_2$, all first derivatives with respect to $\lambda_2$ vanish. Hence, we approximate their second derivatives by taking a second central difference using $\lambda_2=\pm 0.1$ and 0. The total energy is the sum of several large terms with opposing signs. Most contributions are decreasing functions of $\lambda_2$, with the exception of the Ewald energy of repulsion among the positively charged ions which increases due to the short second neighbor bonds.  The repulsion is stronger for Ta than for Hf. Among the negative contributions, the band energy stands out as being stronger for Hf than for Ta. Notice the relative signs of total energy variation, confirming the instability of Hf and the stability of Ta.

\begin{table}
\caption{\label{tab:epartp} Energy contributions $E_0$ to BCC Hf and Ta, and their second variation as $\lambda_2$ varies from -0.1 to 0.1. $\alpha Z$ and $E_{Ewald}$ give the electrostatic energy of the ions in the electron gas. $V_H$ is the Hartree potential. $E_{xc}-V_{xc}$ and $PAW_{dc}$ are double counting corrections. $E_{band}$ is the sum of Kohn-Sham eigenvalues, and $E_{atom}$ is an arbitrary offset approximating the energy of an isolated atom. Units are eV/atom.}
\begin{tabular}{r|rr|rr}
\toprule
                   &   \multicolumn{2}{c|}{Hf}  &   \multicolumn{2}{c}{Ta}  \\
Contribution       &   $E_0$ &~$\Delta^2 E/\Delta\lambda_2^2$ 
                   &   $E_0$ &~$\Delta^2 E/\Delta\lambda_2^2$ \\
\hline                      
$\alpha Z$         &   86.33  &  -9.46    &  125.99  &  -13.20  \\
$E_{Ewald}$        &~-742.93  & +29.64    &~-956.72  &  +36.59  \\
$-V_H$             & -104.39  &  -8.50    & -112.11  &  -13.37  \\
$E_{xc}-V_{xc}$    &   20.13  &  -0.47    &   22.79  &   -0.47  \\
$PAW_{dc}$         &   11.14  &  -0.42    &   13.63  &   -0.06  \\
$E_{band}$         & -140.89  & -11.79    & -152.54  &   -8.67  \\
$E_{atom}$         &  860.87  &   0       &~1047.14  &    0     \\
\hline                      
$E_{Total}$        &   -9.72  &  -0.98    &  -11.80  &   +0.78  \\
\end{tabular}
\end{table}

\subsubsection{Band Structure}

\begin{figure}
\includegraphics[width=3.0in]{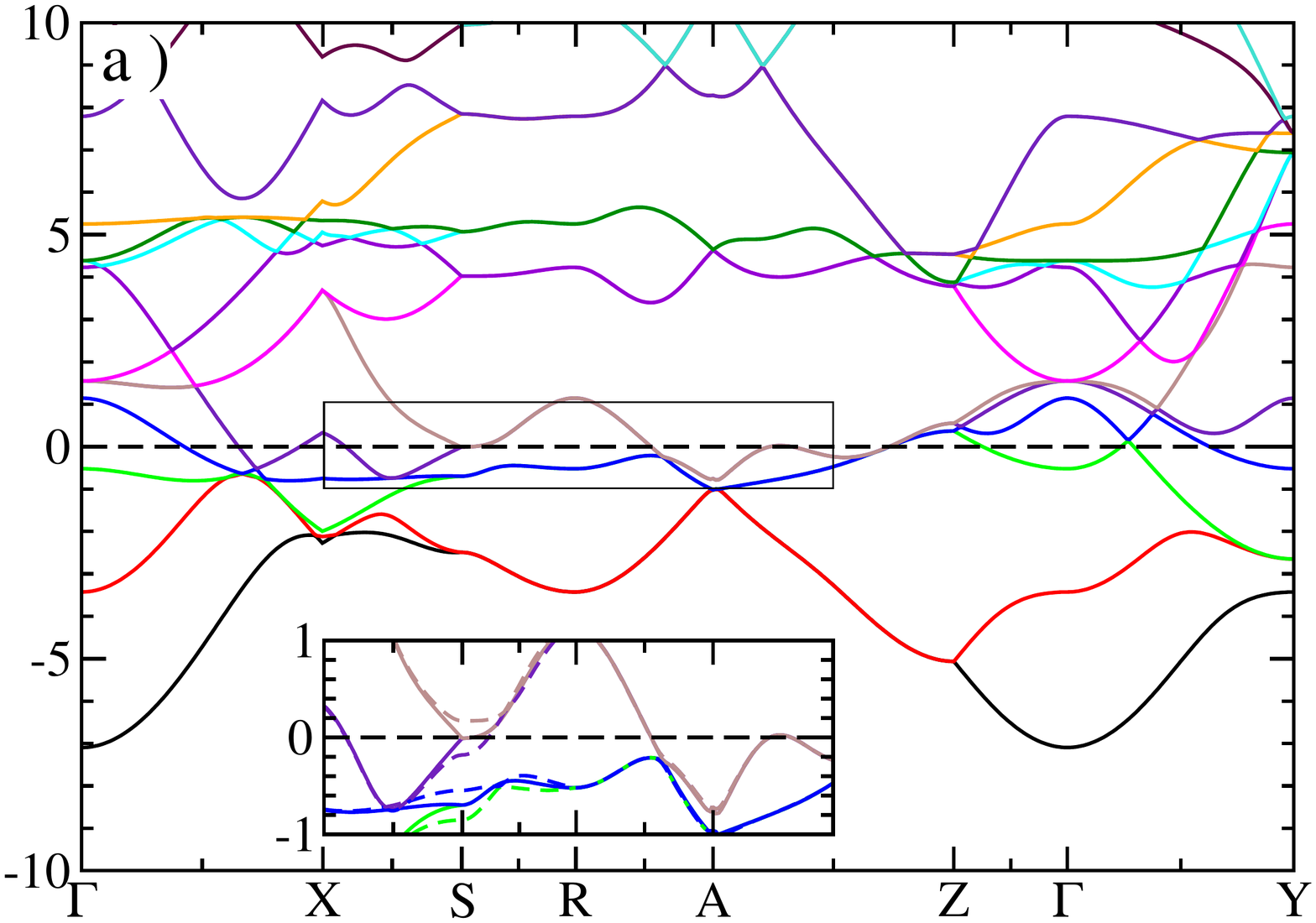}
\includegraphics[width=3.0in]{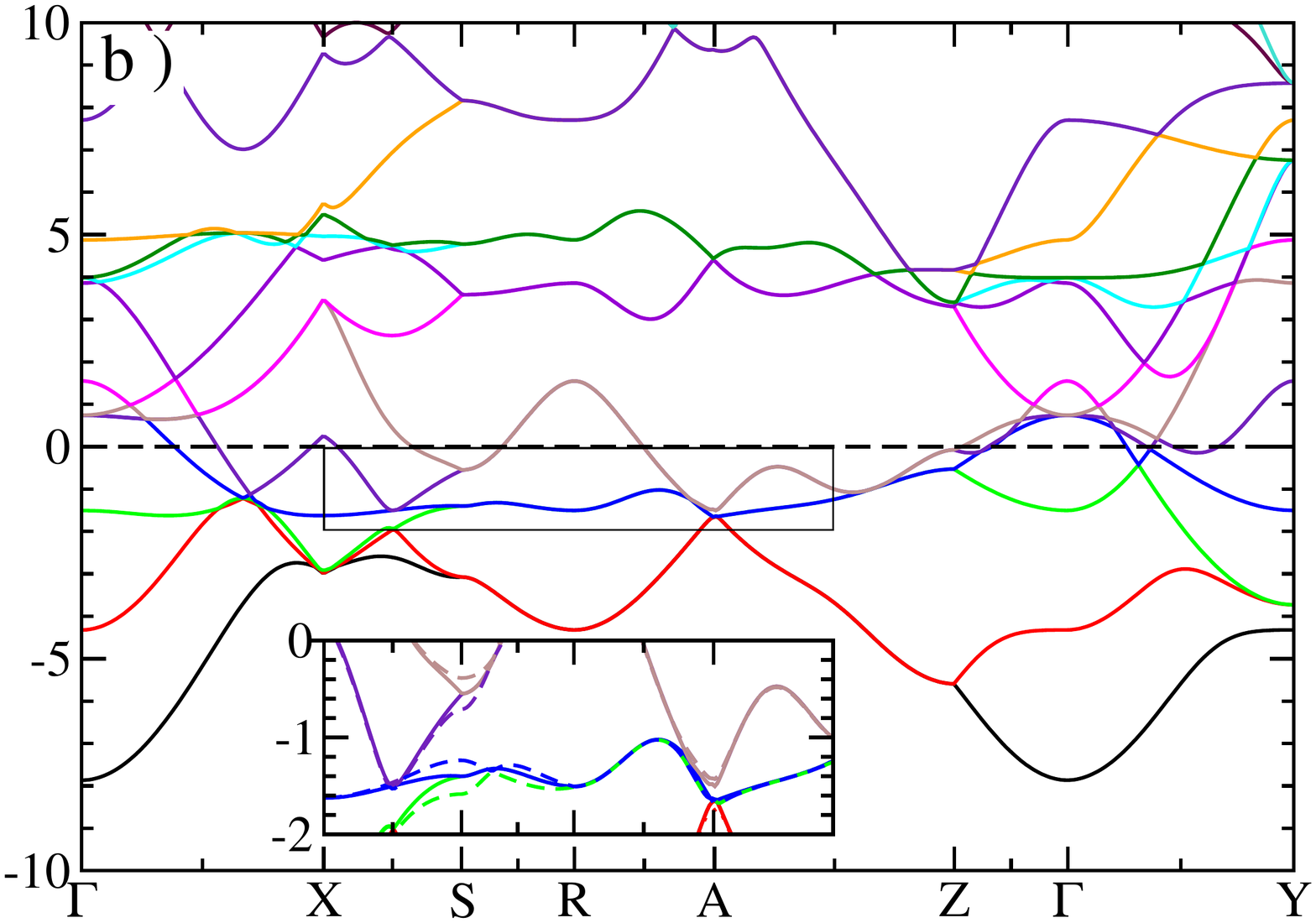}
\caption{Band structure comparison of (a) Hf and (b) Ta. See Fig.~\ref{fig:BZ} and Table~\ref{tab:path} for special point locations. The insets enlarge the vicinity of the special point S. Solid bands show $\lambda_2=0$ ({\em i.e.} BCC in an oS4 setting) while dashed bands (see inset) show $\lambda_2=0.1$.}
\label{fig:bandp}
\end{figure}

\begin{figure}
\includegraphics[width=4in]{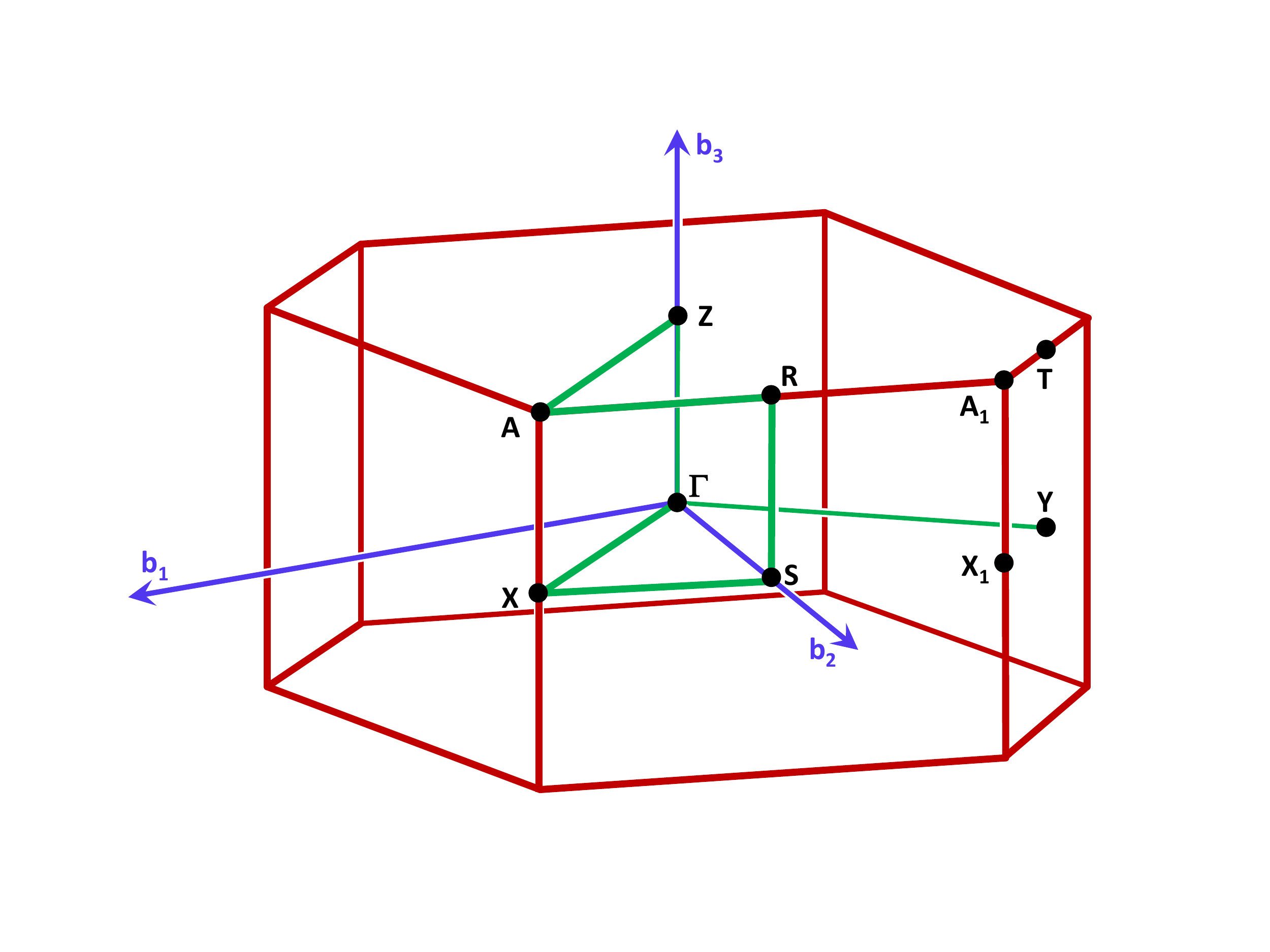}
\caption{\label{fig:BZ} Brillouin zone of oS4 with $b=c=\sqrt{2}a$. See Table~\ref{tab:path} for coordinates of special points.}
\end{figure}

Pseudogap opening in the DOS results from a gap opening in the band structure. Fig.~\ref{fig:bandp} plots the band structures of Hf and Ta at $\lambda_2$=0 and 0.2. These are calculated using the oS4 primitive cell. Fig.~\ref{fig:BZ} displays the Brillouin zone of the BCC structure in the oS4 setting with $\lambda_2=0$, and table~\ref{tab:path} gives coordinates of the special points~\cite{setyawan2010high}. Because oS4 is a supercell of BCC, the usual BCC Brillouin zone is folded. The BCC special points map onto special points of oS4, so that the BCC 4x point H appears at the oS4 point Y; the BCC 2x point N appears at the oS4 points $\Gamma$, R and Y; the BCC 3x point P appears at a position 2/3 of the way along the oS4 special line $\Gamma$X.

As $\lambda_2$ increases, a band-gap opens up between degenerate states at the S point, reflecting the DOS pseudo-gap opening both in Hf and in Ta. For Hf, the gap opens at $E_F$ so that occupied states drop in energy while empty states rise. In contrast, the extra electron in Ta places $E_F$ above the gap so that the drop in energy is partially offset by the increase in energy of some occupied states. Hence $\lambda_2$ has a greater influence on band energy for Hf than for Ta.

Since $\lambda_2$ reduces the symmetry from cubic to orthorhombic, we recognize the the energy reduction by gap opening as a bulk crystalline analogue of the Jahn-Teller distortion. According to Jahn and Teller~\cite{Jahn1937}, breaking the symmetry of a molecule can split a partially occupied highest molecular orbital (HOMO), resulting in a drop in energy of the HOMO and increase in energy of the split-off lowest unoccupied molecular orbital (LUMO). Equivalently, from the point of view of Peierls~\cite{Peierls1955}, symmetry breaking creates a gap in a partially filled band, reducing the energy of occupied states and increasing the energy of unoccupied states. Hence we recognize the mechanical instability of BCC Hf as a manifestation of the Jahn-Teller-Peierls mechanism.

\begin{table}
\caption{\label{tab:path} Symmetry k-points of the oS4 cell~\cite{setyawan2010high}. $\zeta=(1+a^2/b^2)/4=\frac{3}{8}$ for the case $a=b$.}
\begin{tabular}{cccc|cccc}
\toprule
$\times\bb_1$ & $\times\bb_2$ & $\times\bb_3$ &  & 
$\times\bb_1$ & $\times\bb_2$ & $\times\bb_3$ &   \\
\hline
0        & 0         & 0   & $\Gamma$  & -1/2     & 1/2       & 1/2 & T \\
$\zeta$  & $\zeta$   & 1/2 & A         & $\zeta$  & $\zeta$   & 0   & X \\
$-\zeta$ & $1-\zeta$ & 1/2 & A$_1$     & $-\zeta$ & $1-\zeta$ & 0   & X$_1$ \\
0        & 1/2       & 1/2 & R         & -1/2     & 1/2       & 0   & Y \\
0        & 1/2       & 0  & S          & 0        & 0         & 1/2 & Z \\
\end{tabular}
\end{table}

We examined all the elements in the Sc-Cr columns of the periodic table. The entire Ti column shows the same two band degeneracies at the S-point that is lifted by $\lambda_2$  as seen in Hf. These degenerate points fall below $E_F$ throughout the entire V and Cr columns which contain, respectively, one and two electrons more than the Ti column. For the Ti column, the upper degenerate point sits just at $E_F$. In the case of the trivalent elements of the Sc column, which contain one electron less than the Ti column, $E_F$ lies just slightly above the lower degenerate point. In view of their mechanical instability we suspect that proximity to the degenerate point is sufficient to drive the BCC to HCP distortion. Thus we expect similar behavior across the rare earth series, all of which can be trivalent and all of which exhibit transitions from BCC at high temperature to either HCP or FCC.

\subsubsection{Wave Function and Charge Density}

\begin{figure}
\includegraphics[width=4in]{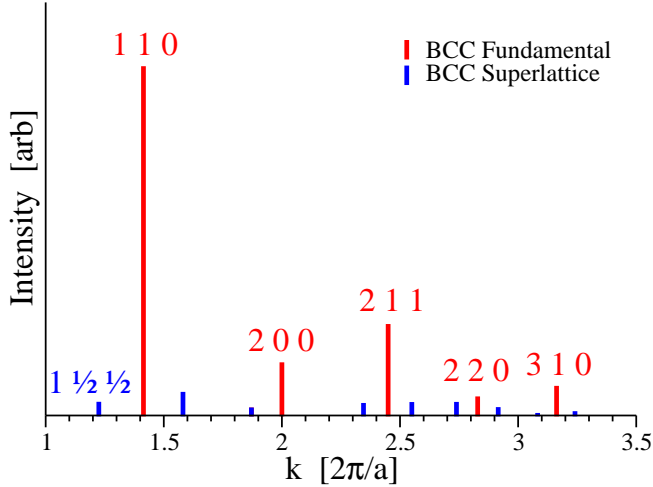}
\caption{\label{fig:dp} Diffraction pattern of oS4 with $\lambda_2=0.5$. With $\lambda_2=0$ only BCC peaks (red) are present, while superlattice peaks (blue) grow linearly in $\lambda_2$. Peaks are indexed according to their positions within the conventional cubic unit cell.}
\end{figure}

With $\lambda_2=0$, the oS4 structure shown in Fig.~\ref{fig:burgers}a becomes BCC, as can be verified from its diffraction pattern illustrated in Fig.~\ref{fig:dp}. As $\lambda_2$ grows, superlattice peaks appear and grow in amplitude proportionally to $\lambda_2$. In real space the structure evolves an alternating pattern of short and long next nearest neighbor bonds (see Fig.~\ref{fig:burgers}b). Notice that this alternation doubles the periodicity along the $y$-axis of the charge density integrated over x and z. The peak at the lowest $k$ value arises at position $\textbf{G}=\bb_2=2\bk_S$ in the notation of oS4 (see Table~\ref{tab:path}) where $\bk_S$ is the position of the S point. This oS4 reciprocal lattice vector is equivalent to the super lattice peak $(hkl=1\frac{1}{2}\frac{1}{2})$ in conventional cubic unit cell indexing.

\begin{figure}
\includegraphics[trim=40mm 40mm 40mm 40mm, clip, width=3in]{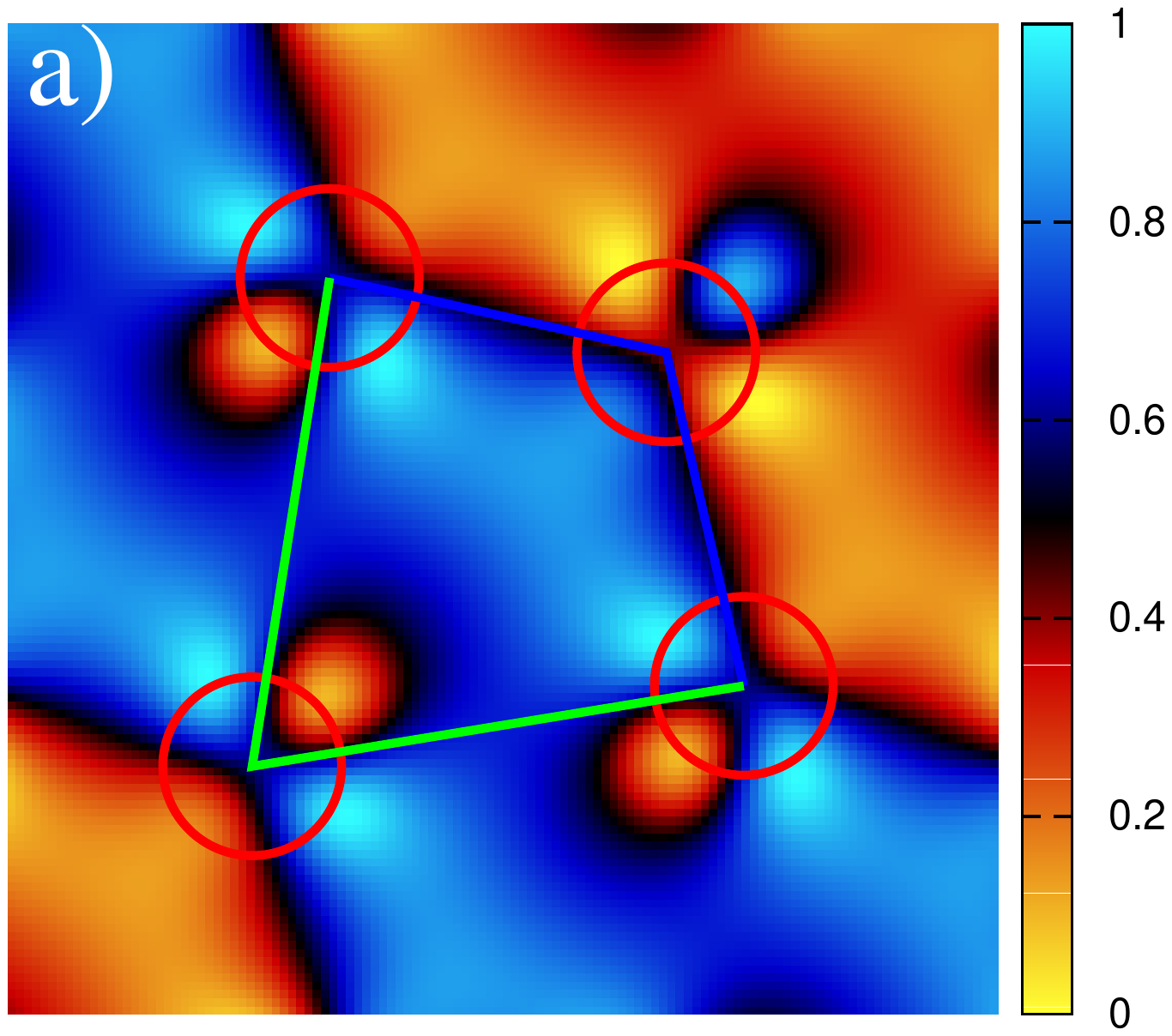}
\includegraphics[trim=40mm 40mm 40mm 40mm, clip, width=3in]{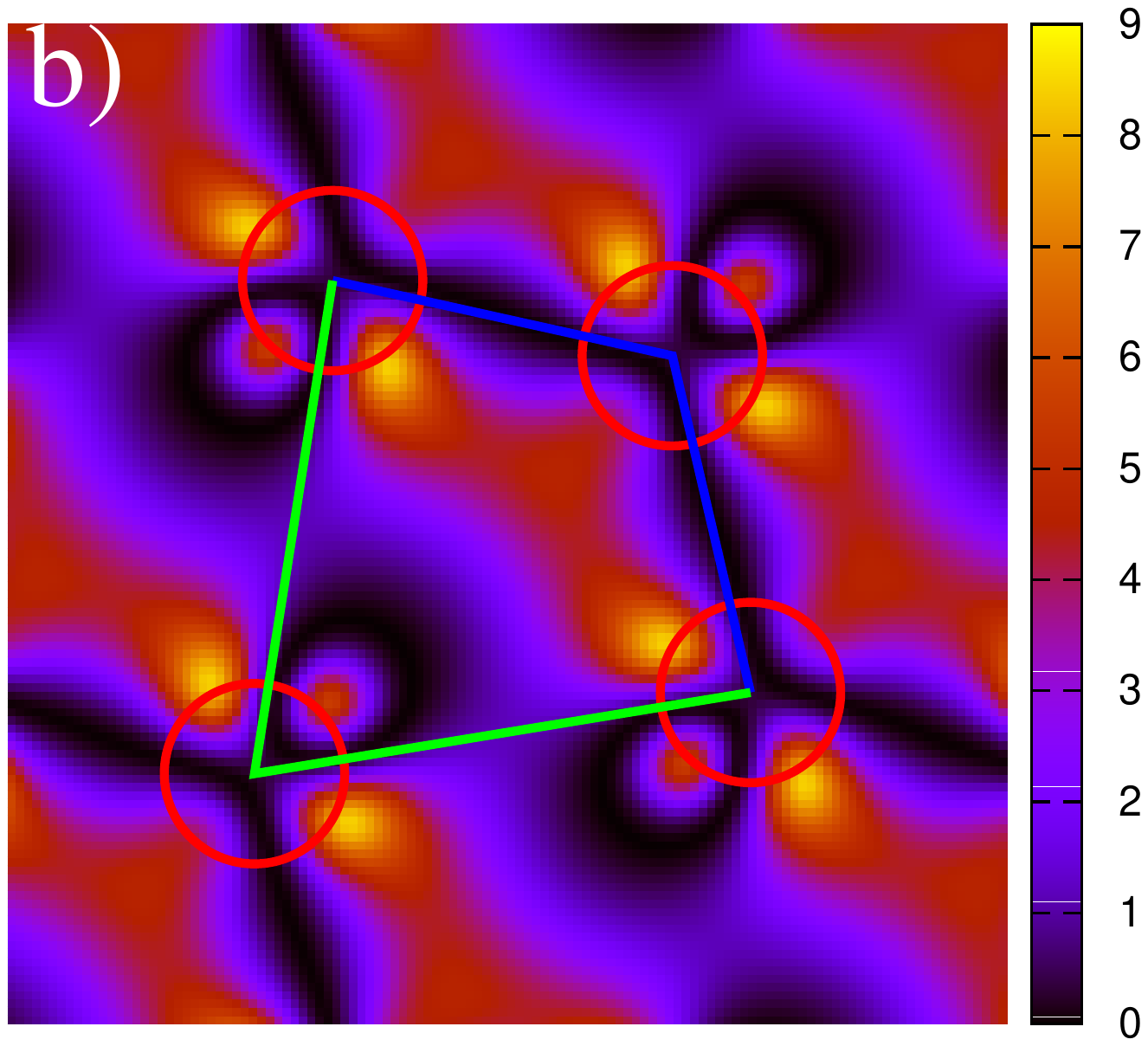}
\includegraphics[trim=40mm 40mm 40mm 40mm, clip, width=3in]{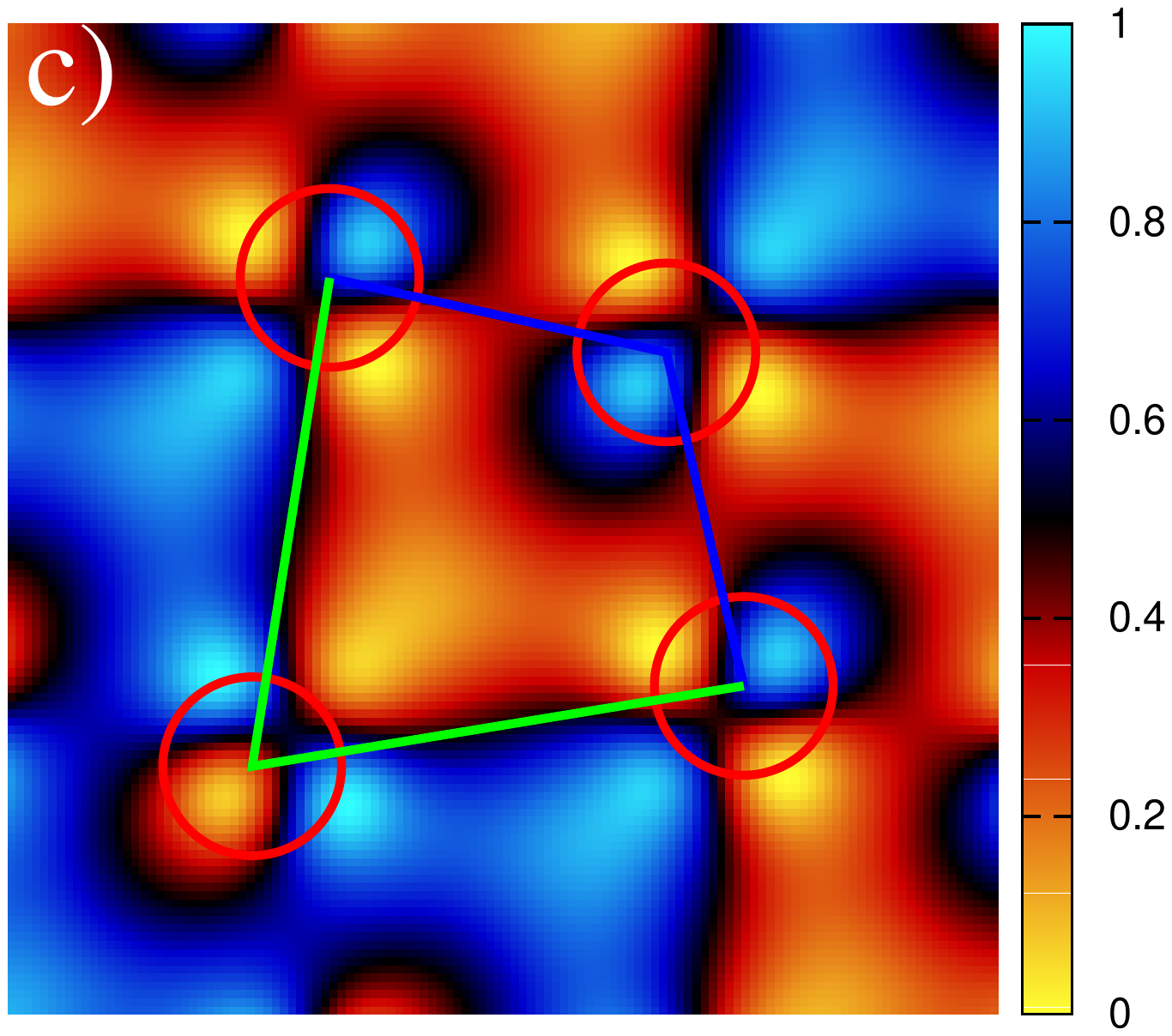}
\includegraphics[trim=40mm 40mm 40mm 40mm, clip, width=3in]{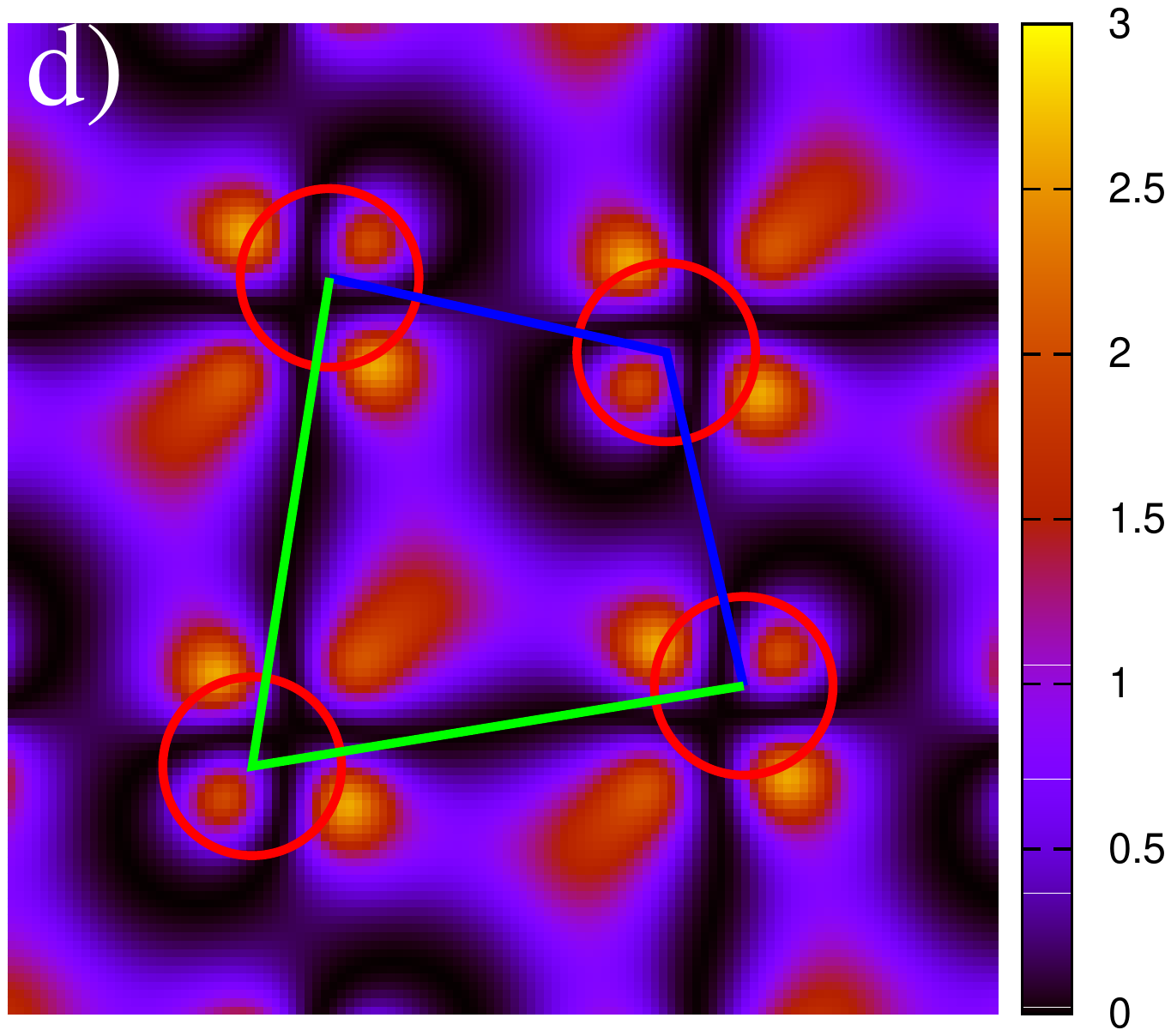}
\caption{\label{fig:wf} S-point wavefunctions of Hf with $\lambda_2=0.5$. The figures show a $2\times 2$ supercell of the conventional cubic unit cell in the same orientation as Fig.~\ref{fig:modes}. a) and b) show the occupied state while c) and d) show the empty state. a) and c) show the real functions $\psi_{\bk_S}(\br)$; b) and d) show the electron density $|\psi_{\bk_S}(\br)|^2$. The wavefunctions were obtained from VASP using WaveTrans~\cite{WaveTrans}. The red circles represent the position of atoms, and blue/green lines are bonding of the shortened/enlongated NNN bonds.}
\end{figure}

The modulation of the potential at wavevector $\bb_2$ couples the degenerate electron states $\psi_{\bk}(\br)=e^{i\bk\cdot\br}u_{\bk}(\br)$ of wave vectors $\bk=\bk_S$ and $-\bk_S$ to first order in perturbation theory, leading to standing wave states that can localize in regions of low potential in the vicinity of the short NNN bonds, and thereby reducing their energy to first order in $\lambda_2$. Short NNN bonds are strengthened ({\em i.e.} are more electron dense) and long NNN bonds are weakened. Figure~\ref{fig:wf} plots the wavefunctions of the occupied and empty states that split off from $E_F$ at the S-point ({\em i.e.} the dashed brown and indigo lines in Fig.~\ref{fig:bandp}a) in the $yz$-plane passing through an atomic layer. $\psi_{\bk}$ turn out to be real functions at the special $k$-point. This plot reveals that they have $d_{yz}$ character in the vicinity of the atoms.

For the occupied state (dashed purple in Fig.~\ref{fig:bandp}a), the sign of the wavefunction (Fig.~\ref{fig:wf}a) alternates between atoms connected by short NNN$_S$ bonds, so that the signs of the lobes of adjacent $d_{xy}$ orbitals match, creating bonding states with high electron density (Fig.~\ref{fig:wf}b) adjacent to the bonds between the atoms. In contrast, the sign of the wavefunction does not reverse along NNN$_L$ bonds, causing the signs of the lobes of the $d_{xy}$ orbitals to conflict, leading to low electron density between the atoms. The higher energy unoccupied state (dashed brown in Fig.~\ref{fig:bandp}a) exhibits the opposite behavior (Fig.~\ref{fig:wf}c, d), with charge density concentrating adjacent to the long NNN$_L$ bonds. The entire effect is a three-dimensional version of the classical Peierls transition~\cite{Kittel}.


A similar effect is observed in Ta (not shown), because setting $\lambda_2\ne 0$ necessarily creates a superlattice. However, in the case of Ta, both the upper and lower states of broken degeneracy remain occupied, so the impact on band structure energy is reduced.

\section{Binary Alloys}
\subsubsection{Elasticity, Phonons and Energy Landscape}
Given the instability of the BCC/HCP elements and the stability of the normal BCC elements, it is interesting to examine alloys containing both BCC/HCP and BCC elements. In this section, we discuss binary alloys taking the Pearson type cP2 structure with BCC/HCP elements at cube vertices and normal BCC elements at body centers.

Elasticity and phonon calculations for TiV, NbZr and HfTa binaries are summarized in Table ~\ref{tab:phononb}. The cubic elastic constants obey the Born stability rules; $C_{11}+2C_{12}>0$, $C_{11}>C_{12}$ and $C_{44}>0$. However, there are two unstable phonon modes in the 2x2x2 cell, equivalent to those illustrated for pure elements in Fig.~\ref{fig:modes}. The modes with the upper imaginary frequencies have degeneracy 6, and correspond to the same alternating slide displacement as in the pure elemental case. The modes with the lower imaginary frequencies have degeneracy 3, rather than 6, because only the normal BCC elements displace. Presumably this is because the large HCP/BCC atoms force a large cubic lattice constant, and the smaller normal BCC atoms displace to shorten the next-nearest neighbor bond lengths.

\begin{table}
\caption{\label{tab:phononb} Elastic constants (units of GPa) and unstable phonon frequencies (units of THz) of TiV, NbZr and HfTa alloys in the cP2 structure. The left frequencies correspond to the mode in Fig.~\ref{fig:modes}a, while the right ones corresponds to Fig.~\ref{fig:modes}b}
\begin{tabular}{c|c|c|c|c|c|c|c}
\toprule
Alloy & Frequencies     & $C_{11}$ & C$_{12}$ & C$_{44}$ \\
\hline
TiV  & 3.44$i$, 3.43$i$ & 174 &125  & 31 \\
\hline
NbZr & 1.17$i$, 2.41$i$ & 153 &110  & 19 \\
\hline
HfTa & 2.04$i$, 2.28$i$ & 153 & 136 & 54 \\
\end{tabular}
\end{table}

\begin{figure}
\includegraphics[trim = 5mm 60mm 5mm 60mm, clip, width=3in]{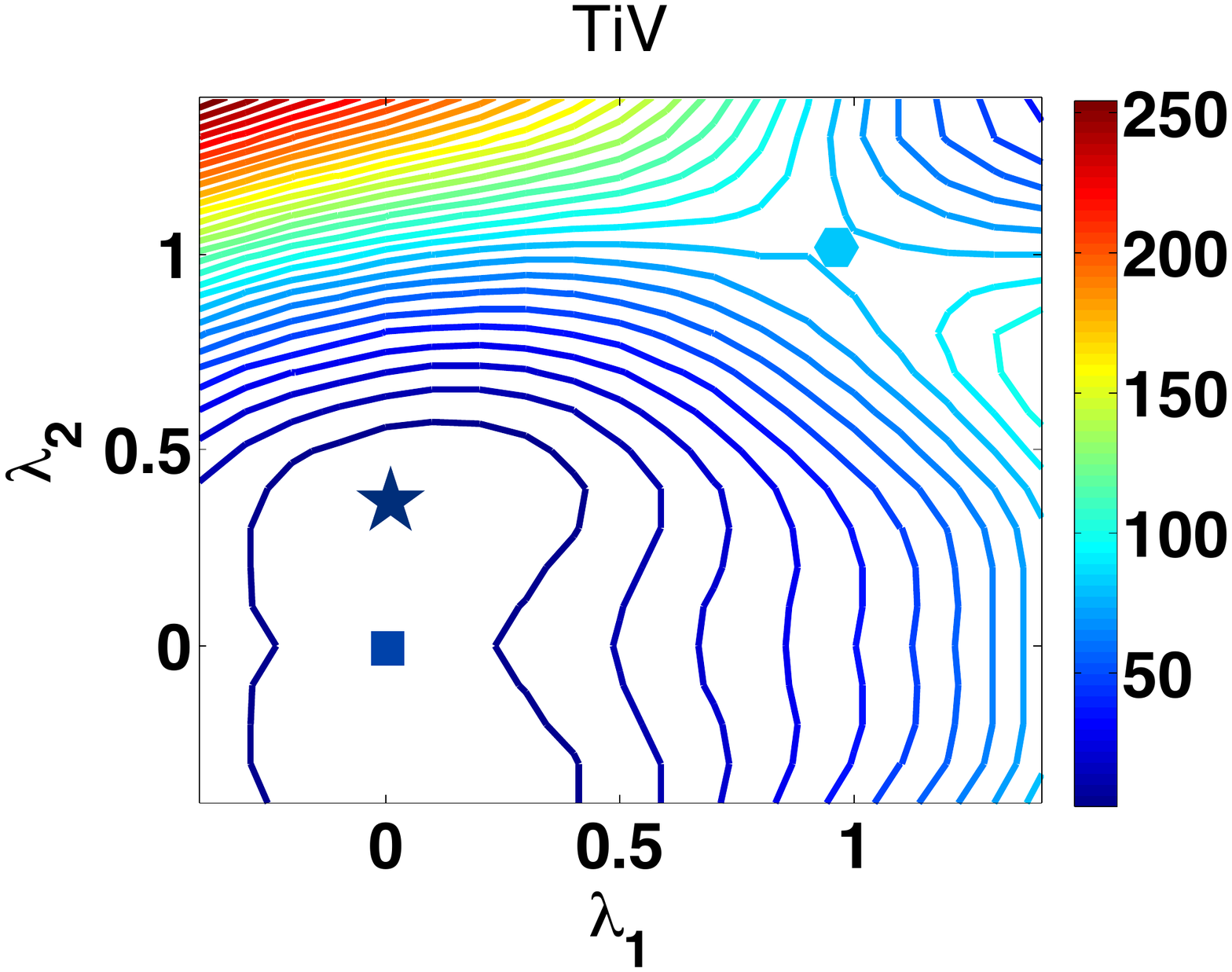}
\includegraphics[trim = 5mm 60mm 5mm 60mm, clip, width=3in]{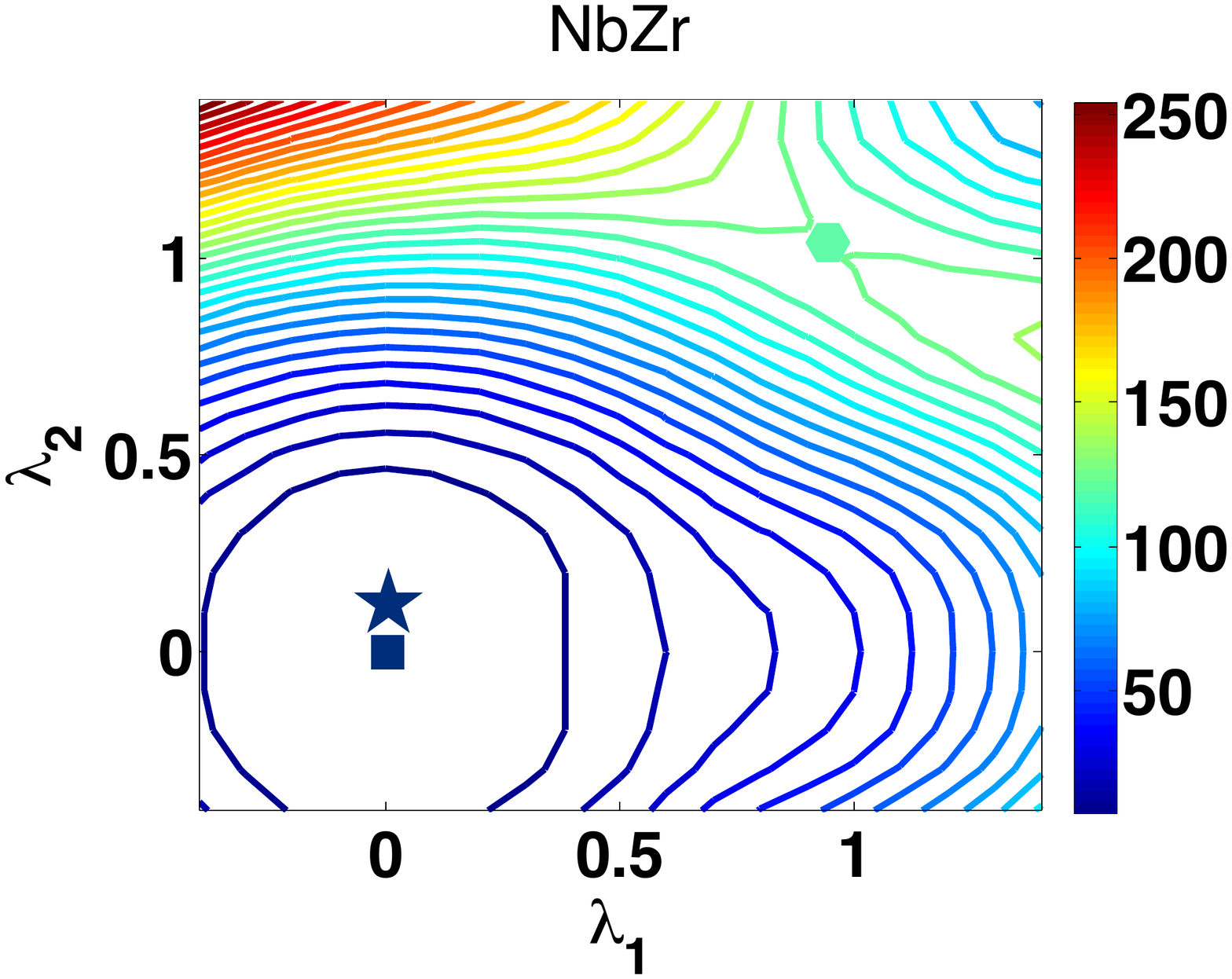}
\includegraphics[trim = 5mm 60mm 5mm 60mm, clip, width=3in]{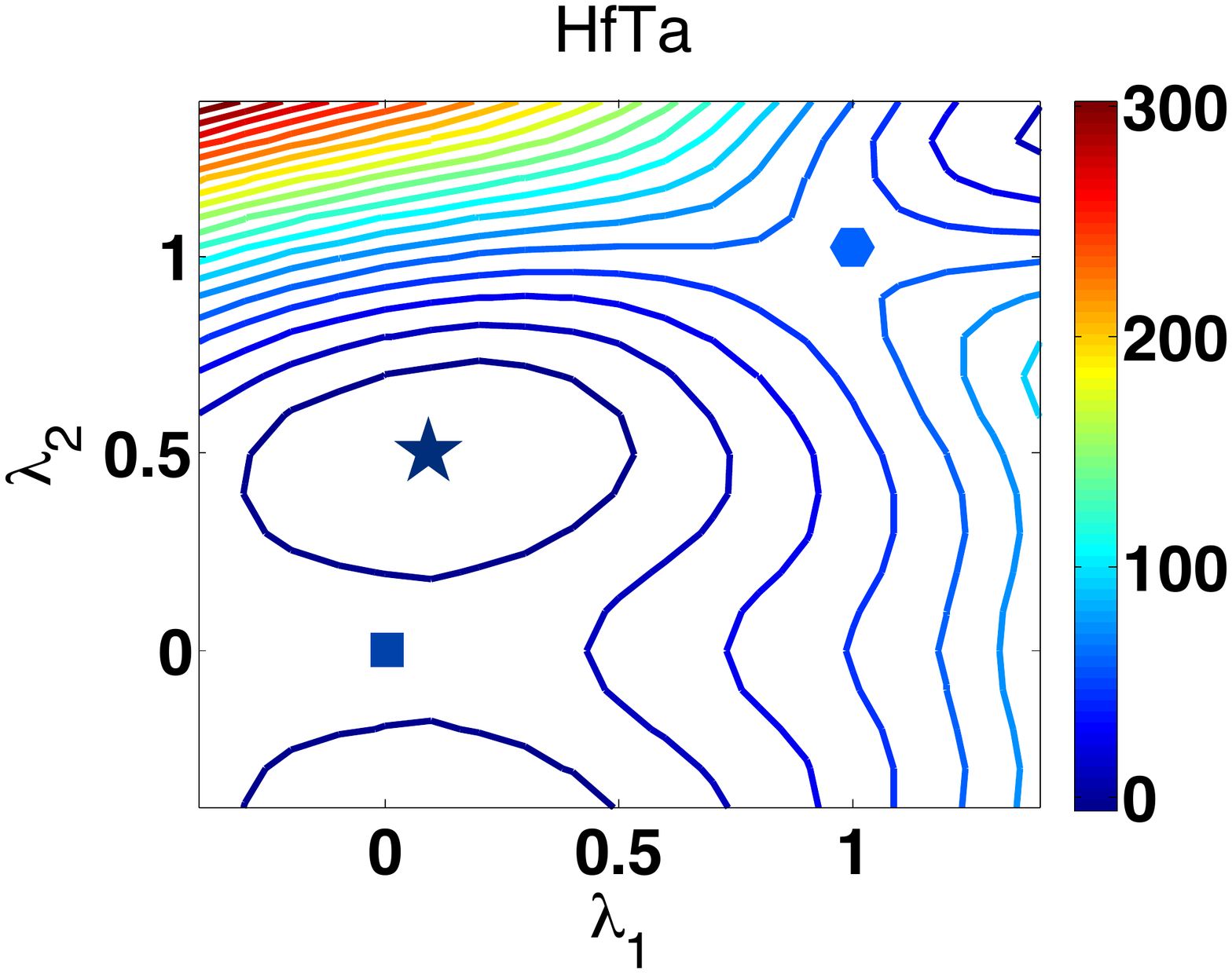}
\caption{\label{fig:elb} Energy landscapes of TiV, NbZr and HfTa (square: cP2, hexagon: oP4, star: most stable state). The colorbars give the relative stability with respect to the BCC structures.}
\end{figure}

Taking the same oS4 structure as in Eqs.~\ref{eq:abc} and ~\ref{eq:R1234}, setting atoms $\bR_1$ and $\bR_2$ to BCC/HCP while  $\bR_3$ and $\bR_4$ to normal BCC, $\lambda_1=\lambda_2=0$ is a cP2 structure, while $\lambda_1=\lambda_2=1$ corresponds to a Pearson type oP4 structure with atoms at HCP positions but with the symmetry reduced to orthorhombic due to the chemical order. Fig.~\ref{fig:elb} shows the energy landscapes of these binary alloys. Unlike the pure BCC/HCP elements and pure BCC elements, neither the BCC nor the HCP structures are stable. Instead, the BCC structures start with the same $\lambda_2$ instability as in the pure BCC/HCP element case, but the transformation gets ``stuck'' part way through. The atoms displace part way along $\lambda_2$ while leaving $\lambda_1$ nearly 0.

\subsection{Electronic Structure}

\begin{figure}
\includegraphics[trim = 8mm 10mm 10mm 30mm, clip, width=4.5in]{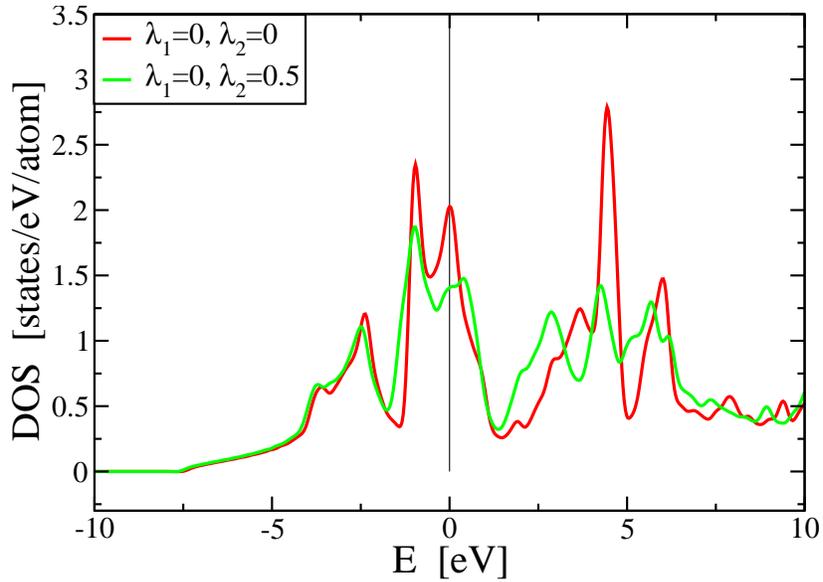}
\caption{\label{fig:dosb} DOS comparison of HfTa before and after the application of $\lambda_2$ distortion.}
\end{figure}

As in Fig.~\ref{fig:dosp} for pure elements, we show the DOS of HfTa before and after the $\lambda_2$ distortion ($\lambda_2$=0.5) in Fig.~\ref{fig:dosb}. Notice that the DOS has a similar shape to pure Hf and Ta, while the Fermi energy lies 0.5eV above the pseudo-gap, compared with 0eV in the case of Hf and 1eV in the case of Ta. As in the case of pure elements, $\lambda_2$ deepens the pseudogap and shifts occupied states to lower energies.

\begin{figure}
\includegraphics[width=4.5in]{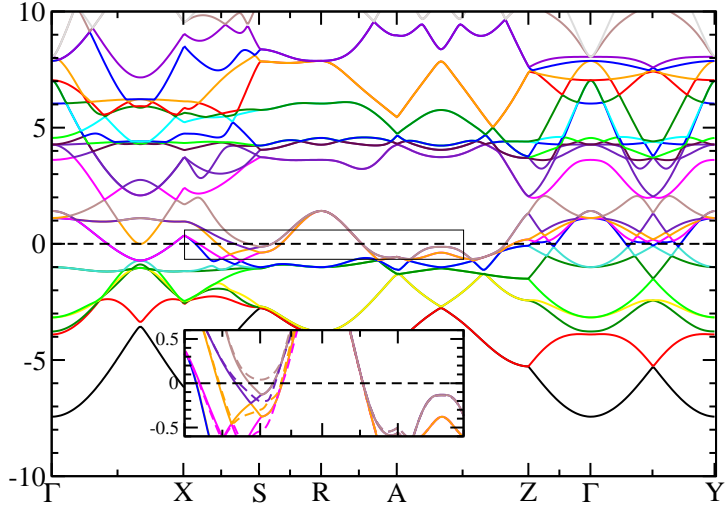}
\caption{Band structure comparison of HfTa before and after the application of $\lambda_2$ distortion.}
\label{fig:bandb}
\end{figure}

Fig.~\ref{fig:bandb} shows the band structure of HfTa before and after the $\lambda_2$ distortion ($\lambda_2$=0.1). At the S point, brown and orange states correspond to the brown state in the pure Hf and Ta cases (Fig.~\ref{fig:bandp}), and indigo and magenta states correspond to the indigo state in Hf and Ta. It worth mentioning that, in the binary alloy cases, in order to achieve Pearson type cP2 structure, a 4-atom unit cell of pearson type oS4 cell is required. Since a 2-atom primitive cell of pearson type oS4 cell is used for pure element cases, the number of bands doubles for binary alloys compared with the pure elements. For binary alloys, before any $\lambda_2$ distortion at S point, these four states already split into two sets of two-fold degenerate states because of the symmetry breaking of the inequivalent atomic sites, but this band gap opening does not stabilize the binary because the gap opening happens below the Fermi energy. After $\lambda_2$ distortion, those two sets of two-fold degenerate states further split because of the further symmetry breaking that both nearest and next nearest neighbor distance become inequivalent like shown in Fig.~\ref{fig:burgers}. This band gap opening makes the purple state below the Fermi energy while the cyan state rises to slightly higher than the Fermi energy, and this accounts for the fact that $\lambda_2$ distortion lowers the net energy of binary HfTa.

Table~\ref{tab:epartb} breaks down the alloy total energy into individual contributions, as was done for pure elements in Table~\ref{tab:epartp}. Again, the electrostatic energy (first three terms) of ionic repulsion stabilizes the BCC structure, while the band energy stands out as a strong destabilizing factor. The net variation $\Delta^2 E/\Delta\lambda_2^2<0$ exhibits instability that is weaker than in the case of pure Hf.

\begin{table}
\caption{\label{tab:epartb} Energy contributions $E_0$ to cP2 HfTa, and their second variation as $\lambda_2$ varies from -0.1 to 0.1. $\alpha Z$ and $E_{Ewald}$ give the electrostatic energy of the ions in the electron gas. $V_H$ is the Hartree potential. $E_{xc}-V_{xc}$ and $PAW_{dc}$ are double counting corrections. $E_{band}$ is the sum of Kohn-Sham eigenvalues, and $E_{atom}$ is an arbitrary offset approximating the energy of an isolated atom. Units are eV/atom.}
\begin{tabular}{r|rr}
\toprule
Contribution       &   $E_0$  & $\Delta^2 E/\Delta\lambda_2^2$   \\
\hline                      
$\alpha Z$         &  105.02  &  -2.50 \\
$E_{Ewald}$        & -847.38  &  +9.74 \\
$-V_H$             & -109.77  &  -4.58 \\
$E_{xc}-V_{xc}$    &   21.47  &  -0.07 \\
$PAW_{dc}$         &   12.51  &  -0.38 \\
$E_{band}$         & -146.52  &  -2.66 \\
$E_{atom}$         &  954.00  &   0    \\
\hline                      
$E_{Total}$        &  -10.66  &  -0.36 \\
\end{tabular}
\end{table}

\begin{figure}
\includegraphics[trim=40mm 40mm 40mm 40mm, clip, width=3in]{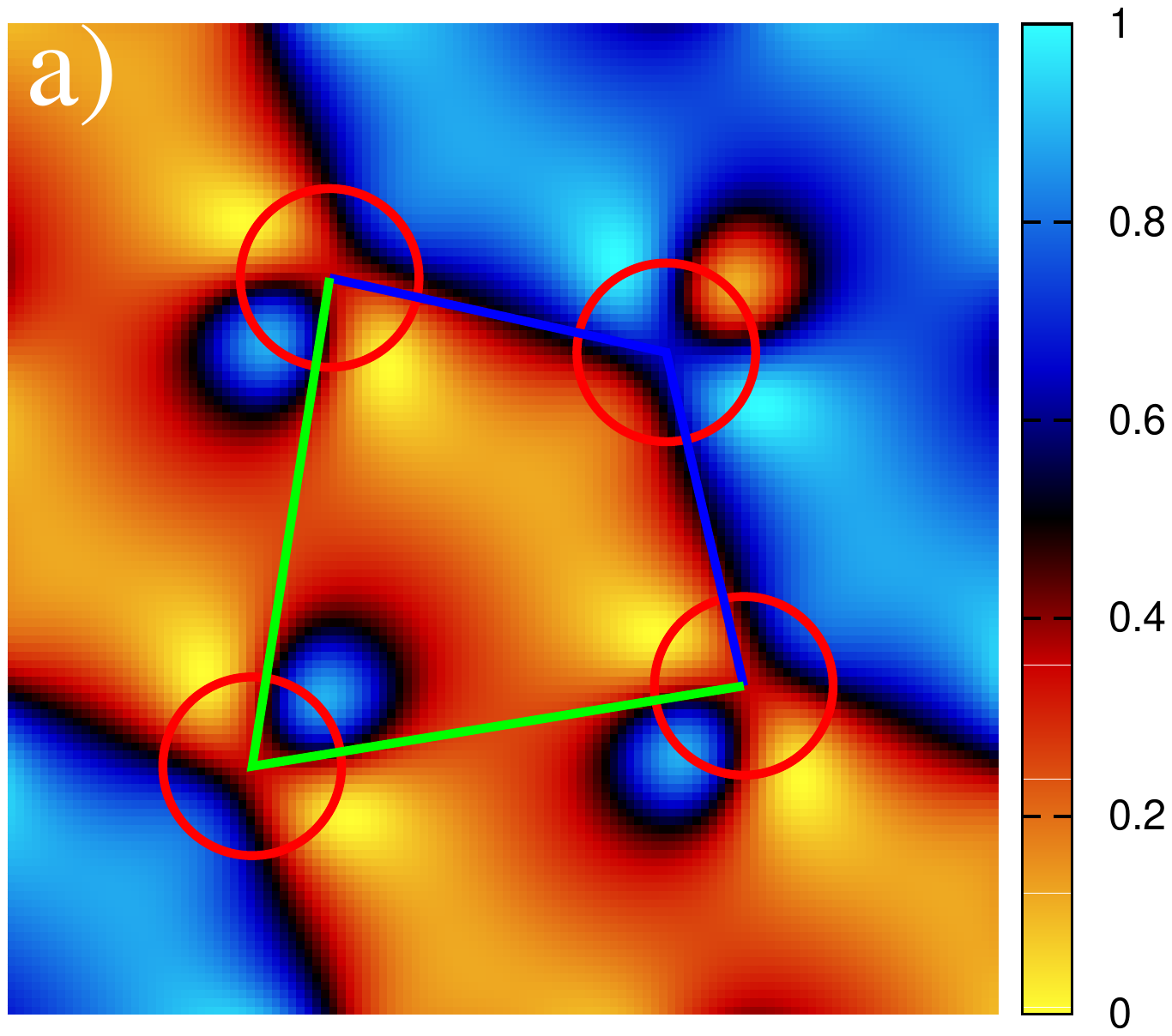}
\includegraphics[trim=40mm 40mm 40mm 40mm, clip, width=3in]{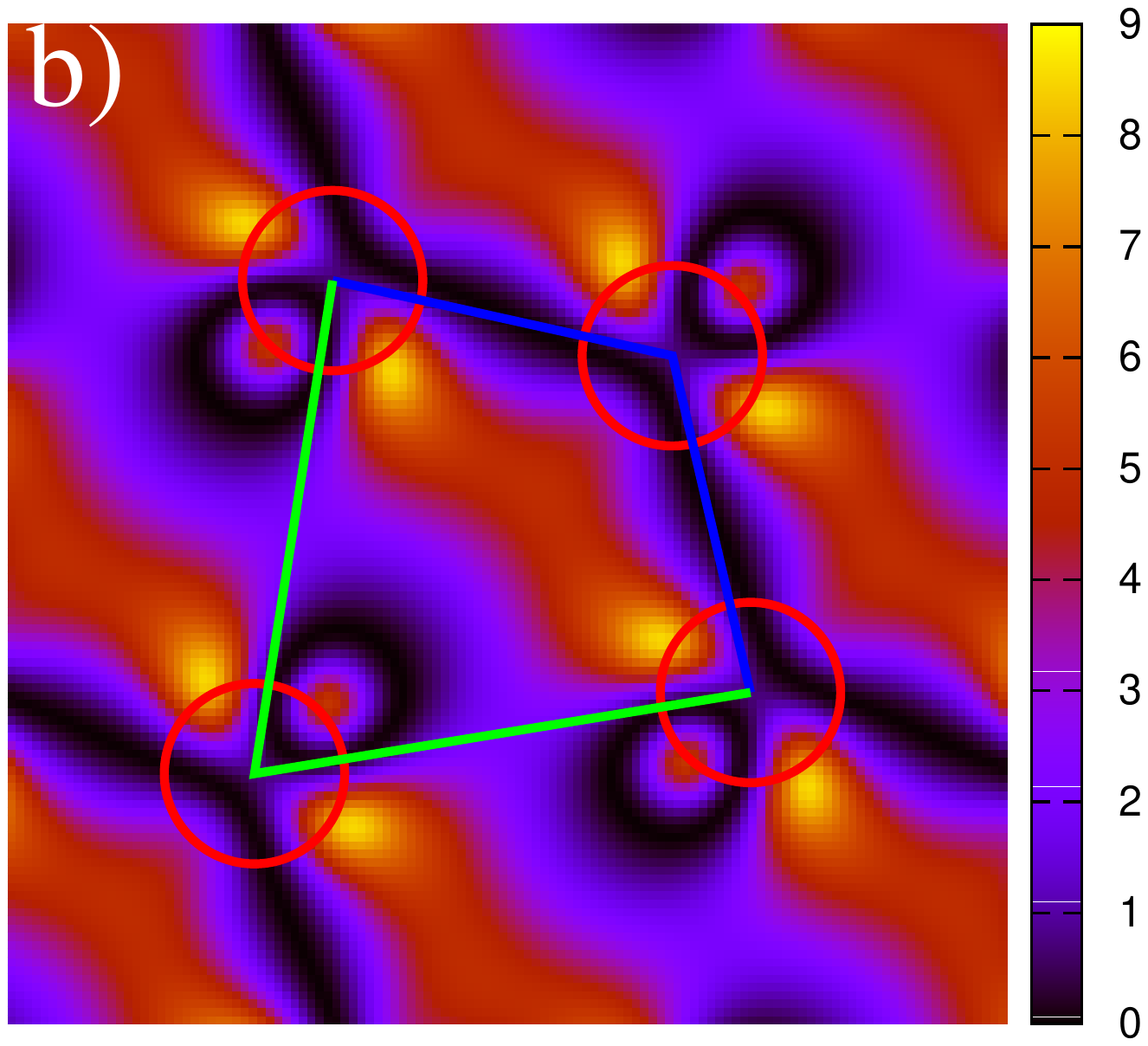}
\includegraphics[trim=40mm 40mm 40mm 40mm, clip, width=3in]{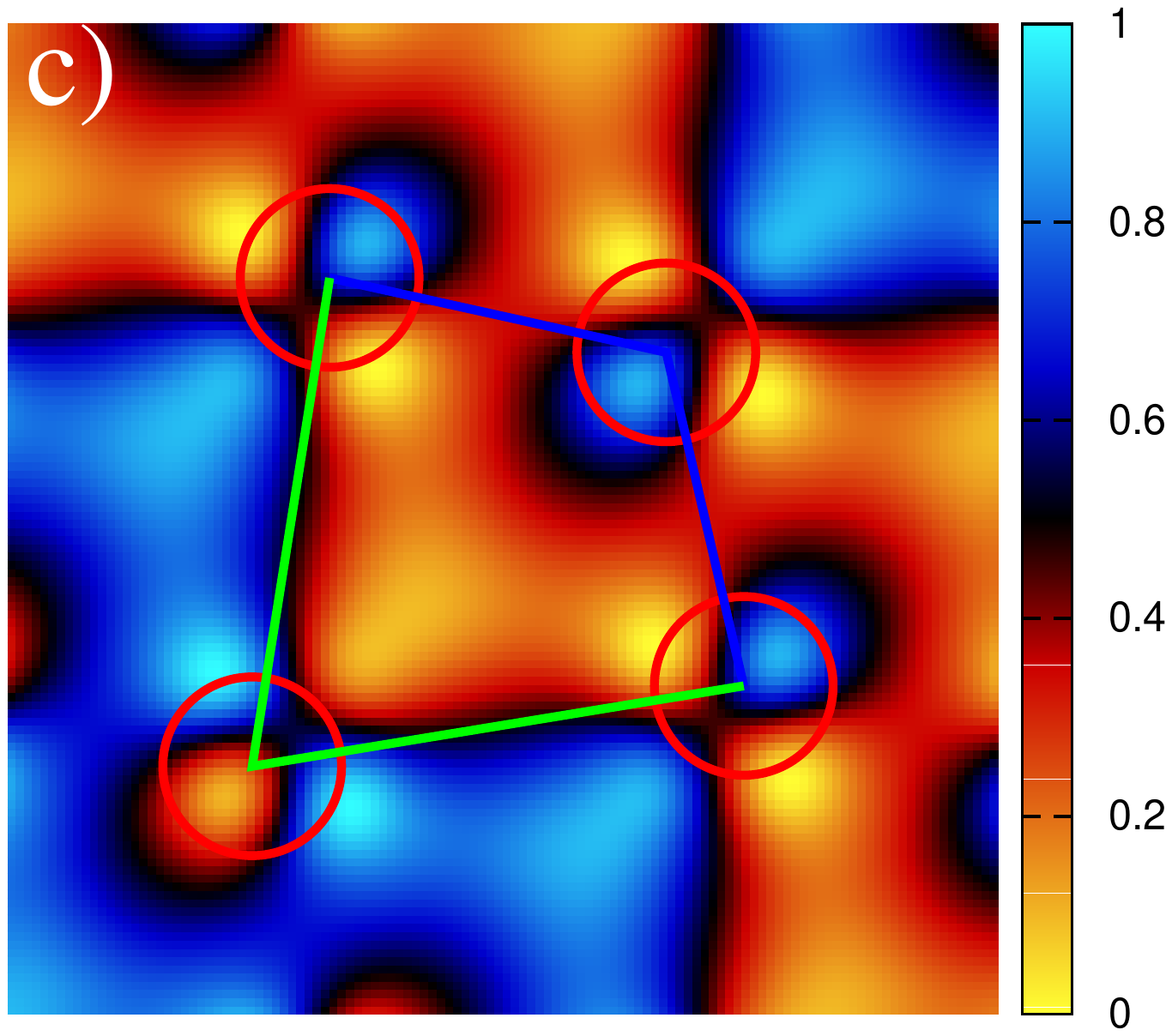}
\includegraphics[trim=40mm 40mm 40mm 40mm, clip, width=3in]{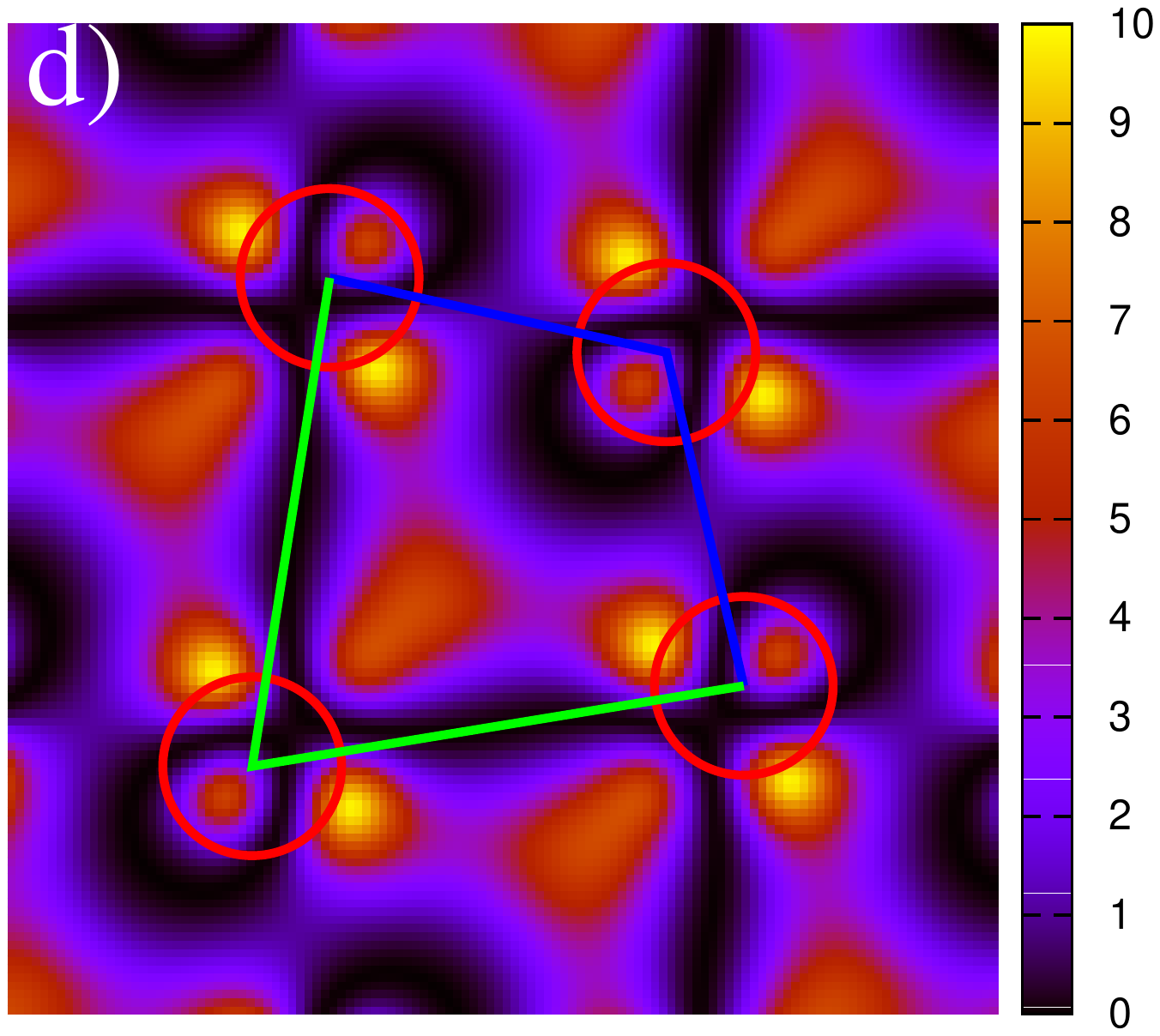}
\caption{\label{fig:wfb} S-point wavefunctions of HfTa with $\lambda_2=0.5$. The figures show a $2\times 2$ supercell of the conventional cubic unit cell in the same orientation as Fig.~\ref{fig:modes}. a) and b) show the occupied state (indigo in Fig.~\ref{fig:bandb}) while c) and d) show the empty state (brown in Fig.~\ref{fig:bandb}). a) and c) show the real function $\psi_{\bk_S}(\br)$; b) and d) show the electron density $|\psi_{\bk_S}(\br)|^2$. The wavefunction was obtained from VASP using WaveTrans~\cite{WaveTrans}. The red circles represent the position of atoms, and blue/green lines are bonding of the shortened/enlongated NNN bonds.}
\end{figure}

Bonding effects in the wave functions at $\bk_S$ are similar to those observed in Hf and Ta, as shown in Fig.~\ref{fig:wfb}.

\section{Conclusions}
This paper describes a complete mechanism of the Burgers distortion of BCC/HCP elements that are stable at high temperatures due to the their vibrational entropy but transition to HCP at low temperatures. The two-stage distortion occurs through an alternating slide displacement between (110) atomic layers followed by relaxation of lattice parameters. The instability is apparent in the violation of elastic stability criteria and the presence of unstable imaginary frequency phonon modes in the BCC state. Electronic structure investigation explains how the distortion lowers the energy: a pseudo-gap in the electronic density of states, a band gap opening at a high symmetry $k$-point, and drop in energy of an occupied bonding state {\em vs.} increased energy of an empty antibonding state.

These effects are similar to the Jahn-Teller instability of molecules that break symmetry to lower the energy of their highest occupied molecular orbital, and are also a three dimensional version of the Peierls instability that creates a superlattice structure in order to open a band gap that lowers the total band energy. They are most striking in tetravalent refractory metals from the Ti column of the periodic table, because in this case the degenerate point sits very close to the Fermi energy. They are also present in the trivalent refractory metals of the Sc column, because in these cases a second degenerate point sits about 0.1 eV below $E_F$. By similar reasoning the effect should be present across the trivalent Lanthanide rare earth series, and we have confirmed this in the case of Lu.

This work does not address the high temperature stability in the BCC state due to vibrational entropy. However, the imaginary modes prevent application of usual techniques for vibrational free energy calculation. Sophisticated techniques are required to incorporate the strong phonon anharmonicity~\cite{Hellman2013,Walle2015} in order to explain stability of BCC at high temperatures.  Normal BCC elements from the V and Cr columns have extra valence electrons so that their Fermi energies sit well above the degenerate points. In these cases, electrostatic ionic repulsion prevents the instability, and their BCC structures are stable at all temperatures.

Finally, we addressed the case of alloys containing both BCC/HCP and normal BCC elements. In this case the instability remains and leads to distortions in atomic positions while leaving lattice constants almost cubic. Perhaps this effect can explain the large lattice distortions reported~\cite{feng2017elastic} in refractory high entropy alloys containing BCC/HCP elements.

\section*{Acknowledgement}
This work was supported by the Department of Energy under grant DE-SC0014506 and by the Pittsburgh Supercomputer Center under XSEDE grant DMR160149.
\section*{References}




\bibliography{ref}

\end{document}